\documentclass[twocolumn,superscriptaddress,amsmath,10pt,floatfix,prd,showpacs,preprintnumbers]{revtex4}
\usepackage[mathscr]{eucal}
\usepackage{color}
\usepackage[dvips]{graphicx}
\usepackage{epsf}
\usepackage{bm}
\newcommand{\be}{\begin{equation}}

\newcommand{\ee}{\end{equation}}
\newcommand{\bea}{\begin{eqnarray}}
\newcommand{\eea}{\end{eqnarray}}

\newcommand{\de}{\partial}

\def\slr#1{\setbox0=\hbox{$#1$}           
   \dimen0=\wd0                                 
   \setbox1=\hbox{/} \dimen1=\wd1               
   \ifdim\dimen0>\dimen1                        
      \rlap{\hbox to \dimen0{\hfil/\hfil}}      
      #1                                        
   \else                                        
      \rlap{\hbox to \dimen1{\hfil$#1$\hfil}}   
      /                                         
   \fi}

\def\be{\begin{eqnarray}}
\def\ee{\end{eqnarray}}

\begin{document}

\title{Testing the Ginzburg-Landau approximation
for three-flavor crystalline color superconductivity}
\date{\today}
\author{Massimo~Mannarelli}
\email{massimo@lns.mit.edu}
\affiliation{Center for Theoretical Physics, Massachusetts Institute
of Technology, Cambridge, MA 02139}
\author{Krishna~Rajagopal}
\email{krishna@lns.mit.edu}
\affiliation{Center for Theoretical Physics, Massachusetts Institute
of Technology, Cambridge, MA 02139}
\affiliation{Nuclear Science Division, MS 70R319,
Lawrence Berkeley National Laboratory, Berkeley, CA 94720}
\author{Rishi~Sharma}
\email{sharma@mit.edu}
\affiliation{Center for Theoretical Physics, Massachusetts Institute
of Technology, Cambridge, MA 02139}

\preprint{MIT-CTP-3725 }

\begin{abstract}
It is an open challenge to analyze the crystalline color superconducting
phases that may arise in cold dense, but not
asymptotically dense, three-flavor quark matter.  At present the only
approximation within which it seems possible to compare the free energies
of the myriad possible crystal structures is the Ginzburg-Landau
approximation.  Here, we test this approximation on a particularly
simple ``crystal" structure in which there are only two condensates
$\langle us \rangle \sim \Delta \exp(i {\bf q_2}\cdot {\bf r})$
and $\langle ud \rangle \sim \Delta \exp(i {\bf q_3}\cdot {\bf r})$
whose position-space dependence is that of two plane waves with
wave vectors ${\bf q_2}$ and ${\bf q_3}$ at arbitrary angles.
For this case, we are able to solve the mean-field gap
equation without making a Ginzburg-Landau approximation.  We find
that the Ginzburg-Landau approximation works in the $\Delta\rightarrow 0$
limit as expected, find that it correctly predicts that
$\Delta$ decreases with increasing angle
between ${\bf q_2}$ and ${\bf q_3}$ meaning that the phase
with ${\bf q_2}\parallel {\bf q_3}$ has the
lowest free energy, and find that
the Ginzburg-Landau approximation is conservative in the
sense that it
underestimates $\Delta$ at all values of the
angle between ${\bf q_2}$ and ${\bf q_3}$.
\end{abstract}
\pacs{12.38.Mh,24.85.+p} \maketitle
\section{Introduction}

Quantum chromodynamics predicts that
at densities that are high enough that baryons are crushed into
quark matter, the quark matter that results features pairing
between quarks at low enough
temperatures, meaning that it is in one of a family of
possible color superconducting phases~\cite{reviews}.
The ``laboratory'' where color superconducting quark matter
is most likely to be found is the interior of compact stars.
Except during the first few seconds after their birth in
supernovae, these stars have temperatures
well below the tens of MeV critical temperature expected
for color superconductivity, meaning that if these stars
feature quark matter cores,  these cores will be color superconductors.

The essence of color superconductivity is quark pairing, leading
to Meissner effects for (some) color magnetic fields, driven by
the BCS mechanism which operates whenever there are attractive
interactions between fermions at a Fermi surface~\cite{BCS}.
The interaction between quarks in QCD is strong and is attractive
between quarks that are antisymmetric in color, so we expect
cold dense quark matter to exhibit color superconductivity.
We shall only consider Cooper pairs whose
pair wave function is antisymmetric in Dirac indices --- the relativistic
generalization of zero total spin. (Other
possibilities have been
investigated~\cite{reviews,Iwasaki:1994ij,Alford:1997zt,Alford:1998mk,Alford:2002kj}
and found to be less favorable.)
This in turn requires antisymmetry  in flavor, meaning in particular
that the two quarks in a Cooper pair must have different flavor.

It is by now well-established that at sufficiently high densities,
where the up, down and strange quarks can be treated
on an equal footing and the disruptive effects of the
strange quark mass can be neglected, quark matter
is in the color-flavor locked (CFL) phase, in which
quarks of all three colors and all three flavors form
conventional Cooper pairs with zero total momentum~\cite{Alford:1998mk,reviews},
and all fermionic
excitations are gapped.   However, even at the very center
of a compact star the quark number chemical potential
$\mu$ cannot be much larger than 500 MeV, meaning
that the strange quark mass $M_s$ (which is density dependent, lying
somewhere between its vacuum current mass of about 100 MeV and constituent
mass of about 500 MeV) cannot be neglected.
Furthermore, bulk matter, as relevant for a compact star, must be in weak equilibrium
and must be electrically and color
neutral~\cite{Iida:2000ha,Amore:2001uf,Alford:2002kj,Steiner:2002gx,Huang:2002zd}.
All these factors work to separate the Fermi momenta of the three different
flavors of quarks, and thus disfavor the cross-species BCS pairing
that characterizes the CFL phase.
If we imagine beginning at asymptotically high densities and reducing
the density, and suppose that CFL pairing is disrupted by the heaviness
of the strange quark before color superconducting quark matter is superseded
by baryonic matter, the CFL phase must be replaced by some
phase of quark matter in which there is less, and less symmetric, pairing.
The nature of this phase is not currently established.

Within a spatially homogeneous ansatz, the next phase down in
density is the gapless CFL (gCFL)
phase~\cite{Alford:2003fq,Alford:2004hz,Alford:2004nf,Ruster:2004eg,Fukushima:2004zq,Alford:2004zr,Abuki:2004zk,Ruster:2005jc}.
In this phase, quarks of all three colors and all three flavors
still form ordinary Cooper pairs, with each pair having zero total
momentum, but there are regions of momentum space in which certain
quarks do not succeed in pairing, and these regions are bounded by
momenta at which certain fermionic quasiparticles are gapless. This
variation on BCS pairing --- in which the same species of fermions
that pair feature gapless quasiparticles --- was first proposed for
two flavor quark matter~\cite{Shovkovy:2003uu} and in an atomic
physics context~\cite{Gubankova:2003uj}.  In all these contexts,
however, the gapless paired state turns out in general to suffer
from a ``magnetic instability'': it can lower its energy by the
formation of counter-propagating
currents~\cite{Huang:2004bg,Casalbuoni:2004tb}. In the atomic
physics context, the resolution of the instability is phase
separation, into macroscopic regions of two phases in one of which
standard BCS pairing occurs and in the other of which no pairing
occurs~\cite{Bedaque:2003hi,KetterleImbalancedSpin,HuletPhaseSeparation}.
In three-flavor quark matter, where the instability of the gCFL
phase has been established in Refs.~\cite{Casalbuoni:2004tb}, phase
coexistence would require coexisting components with opposite color
charges, in addition to opposite electric charges, making it very
unlikely that a phase separated solution can have lower energy than
the gCFL phase~\cite{Alford:2004hz,Alford:2004nf}. Furthermore,
color superconducting phases which are less symmetric than the CFL
phase but still involve only conventional BCS pairing, for example
the much-studied 2SC phase in which only two colors of up and down
quarks pair~\cite{Bailin:1983bm,Alford:1997zt,Rapp:1997zu} but
including also many other possibilities~\cite{Rajagopal:2005dg},
cannot be the resolution of the gCFL
instability~\cite{Alford:2002kj,Rajagopal:2005dg}. It seems likely,
therefore, that a ground state with counter-propagating currents is
required.  This could take the form of a crystalline color
superconductor~\cite{Alford:2000ze,Bowers:2001ip,Casalbuoni:2001gt,Leibovich:2001xr,Kundu:2001tt,Bowers:2002xr,Casalbuoni:2003wh,Casalbuoni:2004wm,Casalbuoni:2005zp,Ciminale:2006sm}
--- the QCD analogue of a form of non-BCS pairing first considered
by Larkin, Ovchinnikov, Fulde and Ferrell~\cite{LOFF}. Or, given
that the CFL phase itself is likely augmented by kaon
condensation~\cite{Bedaque:2001je,Kryjevski:2004jw}, it could take
the form of a phase in which a CFL kaon condensate carries a current
in one direction balanced by a counter-propagating current in the
opposite direction carried by gapless quark
quasiparticles~\cite{Kryjevski:2005qq}. This meson supercurrent
phase has been shown to have a lower free energy than the gCFL
phase. The investigation of crystalline color superconductivity in
three-flavor QCD has only just begun~\cite{Casalbuoni:2005zp}.
Although such phases seem to be free from magnetic
instability~\cite{Ciminale:2006sm}, it remains to be seen whether
such a phase can have a lower free energy than the meson current
phase, making it a possible resolution to the gCFL instability.  The
simplest ``crystal'' structures do not suffice, but experience in
the two-flavor context~\cite{Bowers:2002xr} suggests that realistic
crystal structures constructed from more plane waves will prove to
be qualitatively more robust.  Once we know (or have a
well-motivated conjecture for) the favored crystal structure, the
challenge will then be to calculate its shear modulus and the
pinning force it exerts on rotational vortices trying to move
through it, as these are the microphysical inputs needed to
determine whether observations of pulsar glitches can be used to
rule out (or in) the presence quark matter in the crystalline color
superconducting phase within compact stars \cite{Alford:2000ze}.

Determining the favored crystal structure(s) in the
crystalline color superconducting phase(s) of three-flavor QCD requires
determining the gaps and comparing the free energies for very many candidate structures,
as there are even more possibilities than the many that were investigated in
the two-flavor context~\cite{Bowers:2002xr}.   As there, it seems likely
that progress will require making a Ginzburg-Landau approximation.
This approximation is controlled if $\Delta \ll \Delta_0$,
where $\Delta$ is the gap parameter of the crystalline color superconducting
phase itself and $\Delta_0$ is the gap parameter in the CFL phase that would
occur if $M_s$ were zero.    This approximation will be dubious precisely in
the case of interest: if a crystalline phase exists with a free energy lower
than that of the gCFL phase, such a phase will be characterized by robust
pairing meaning that $\Delta$ will not be much smaller than $\Delta_0$.
Our purpose in this paper is to analyze a particularly simple
one parameter family of ``crystal'' structures
in three-flavor quark matter, simple enough that we can do the analysis
both with and without the Ginzburg-Landau approximation.

We shall work throughout in a Nambu--Jona-Lasinio (NJL)
model in which the QCD interaction between
quarks is replaced by a point-like four-quark interaction, with the quantum
numbers of single-gluon exchange, analyzed in mean field theory.
This is not a controlled approximation.
However, it suffices for our purposes: because this model has  attraction
in the same channels as in QCD, its high density phase is the CFL phase; and, the
Fermi surface splitting effects whose
qualitative consequences we wish to study can all be built
into the model.  Note that we
shall assume throughout that $\Delta_0\ll \mu$.  This weak coupling assumption
means that the pairing is dominated by modes near the Fermi surfaces, justifying
the analysis of the model using High Density Effective
Theory (HDET)~\cite{Hong:1998tn,Nardulli:2002ma}.  Thus,
we  shall be comparing two calculations: one making the
Ginzburg-Landau assumption
that $\Delta\ll \Delta_0 \ll \mu$ and the other allowing for $\Delta \sim \Delta_0 \ll \mu$.
The Ginzburg-Landau calculation can be extended to more complicated
crystal structures than the simple ones that we shall analyze.  We shall
find that the Ginzburg-Landau approximation works when it should and that,
at least for the simple crystal structure we analyze, it is conservative in the
sense that when it breaks down it always underestimates the gap $\Delta$ and the
condensation energy.  Furthermore, we find that the Ginzburg-Landau approximation
correctly determines which crystal structure among the one parameter family
that we analyze has the largest gap and lowest free energy.  We shall see,
however, that the range of validity of the Ginzburg-Landau approximation
does depend on the crystal structure.

Our paper is organized as follows. In Section \ref{general} we shall describe
the one parameter family of three-flavor crystalline color superconductors that
we analyze.  In Section \ref{GLsection} we present the
results of our Ginzburg-Landau analysis. In Section \ref{HDETsection}, we
recast the High Density Effective Theory slightly, as needed for our purposes,
and redo our analysis without the Ginzburg-Landau approximation.
In Section \ref{results} we show the numerical results of our analyses, make
comparisons, and draw conclusions.

\section{Model and Ansatz\label{general}}

We shall analyze quark matter containing massless $u$ and $d$ quarks and $s$ quarks with
an effective mass $M_s$.  (Although the strange quark mass can be
determined self-consistently by solving for an $\langle \bar s s\rangle$
condensate~\cite{Steiner:2002gx,Abuki:2004zk,Ruster:2005jc}, we shall leave this to future
work and treat $M_s$ as a
parameter.) The Lagrangian density describing this system in the absence
of interactions is given by
\begin{equation}
{\cal L}_0=\bar{\psi}_{i\alpha}\,\left(i\,\de\!\!\!
/^{\,\,\alpha\beta}_{\,\,ij} -M_{ij}^{\alpha\beta}+
\mu^{\alpha\beta}_{ij} \,\gamma_0\right)\,\psi_{\beta j}
\label{lagr1}\ \,,
\end{equation}
where $i,j=1,2,3$ are flavor indices and $\alpha,\beta=1,2,3$ are
color indices, where we  have suppressed the Dirac indices,
where $M_{ij}^{\alpha\beta} =\delta^{\alpha\beta}\, {\rm
diag}(0,0,M_s)_{ij} $ is the mass matrix, where
$\de^{\alpha\beta}_{ij}=\partial\delta^{\alpha\beta}\delta_{ij}$ and where
the quark chemical potential matrix is  given by
\begin{equation}\mu^{\alpha\beta}_{ij}=(\mu\delta_{ij}-\mu_e
Q_{ij})\delta^{\alpha\beta} + \delta_{ij} \left(\mu_3
T_3^{\alpha\beta}+\frac{2}{\sqrt 3}\mu_8 T_8^{\alpha\beta}\right) \,
, \label{mu}
\end{equation} with  $Q = {\rm
diag}(2/3,-1/3,-1/3)_{ij} $ the quark electric-charge matrix and
$T_3$ and $T_8$ the Gell-Mann matrices in color space. We shall
quote results at quark number chemical potential $\mu=500$~MeV
throughout.

In QCD, $\mu_e$, $\mu_3$ and $\mu_8$ are the zeroth components of
electromagnetic and color gauge fields, and the gauge field dynamics
ensure that they take on values such that the matter is
neutral~\cite{Alford:2002kj,Gerhold:2003js}, satisfying
\be
\label{neutrality}
\frac{\partial \Omega}{\partial\mu_e} =
\frac{\partial \Omega}{\partial\mu_3} =
\frac{\partial \Omega}{\partial\mu_8} = 0\ ,
\ee
with $\Omega$ the free energy density of the system.
In the NJL model that we shall employ, in which quarks interact
via four-fermion interactions and there are no gauge fields, we introduce
$\mu_e$, $\mu_3$ and $\mu_8$ by hand, and choose them to satisfy
the neutrality constraints (\ref{neutrality}).  The assumption of weak equilibrium
is built into the calculation via the fact that the only flavor-dependent chemical
potential is $\mu_e$, ensuring for example that the chemical potentials of
$d$ and $s$ quarks with the same color must be equal.  Because
the strange quarks have greater mass, the equality
of their chemical potentials implies that the $s$ quarks  have smaller
Fermi momenta than the $d$ quarks in the absence
of BCS pairing.  In the absence of pairing, then, because weak equilibrium
drives the massive strange quarks to be less numerous than
the down quarks, electrical neutrality
requires a $\mu_e>0$, which makes the up quarks less numerous than the
down quarks and introduces some electrons into the system.
In the absence of pairing, color neutrality is obtained with $\mu_3=\mu_8=0.$

The Fermi momenta of the quarks and electrons in quark matter
that is electrically and color neutral and in weak equilibrium
are given in the absence of pairing by
\begin{eqnarray}
p_F^d &=& \mu+\frac{\mu_e}{3}\nonumber\\
p_F^u &=& \mu-\frac{2 \mu_e}{3}\nonumber\\
p_F^s &=& \sqrt{\left(\mu+\frac{\mu_e}{3}\right)^2- M_s^2} \approx \mu + \frac{\mu_e}{3}
-\frac{M_s^2}{2\mu}\nonumber\\
p_F^e &=& \mu_e\ ,
\label{pF1}
\end{eqnarray}
where we have simplified $p_F^s$ upon assuming that $M_s$ and $\mu_e$ are
small compared to $\mu$ by working only to linear order in $\mu_e$ and $M_s^2$.
To this order, electric neutrality requires
\be
\mu_e=\frac{M_s^2}{4\mu}\ ,
\ee
yielding
\begin{eqnarray}
p_F^d &=& \mu+\frac{M_s^2}{12\mu}=p_F^u+\frac{M_s^2}{4\mu}\nonumber\\
p_F^u &=& \mu-\frac{M_s^2}{6\mu}\nonumber\\
p_F^s &=& \mu-\frac{5 M_s^2}{12 \mu} =p_F^u-\frac{M_s^2}{4\mu}\nonumber\\
p_F^e &=& \frac{M_s^2}{4\mu}\ .
\label{pF2}
\end{eqnarray}
We see from (\ref{pF1}) that to leading order in $M_s^2$ and $\mu_e$, the
effect of the strange quark mass on unpaired quark matter is as if instead
one reduced the strange quark chemical potential by $M_s^2/(2\mu)$.
We shall make this approximation throughout.
The corrections to this approximation in
an NJL analysis of a two-flavor crystalline
color superconductor have been evaluated and found
to be small~\cite{Kundu:2001tt}, and we expect the same to be true here.
Upon making this assumption, we need no longer be careful about the
distinction between $p_F$'s and $\mu$'s, as we can simply think of the three
flavors of quarks as if they have chemical potentials
\begin{eqnarray}
\mu_d &=& \mu_u + 2 \delta\mu_3 \nonumber\\
\mu_u &=&p_F^u \nonumber\\
\mu_s &=& \mu_u - 2 \delta\mu_2
\label{pF3}
\end{eqnarray}
with
\be
\delta\mu_3 = \delta\mu_2 = \frac{M_s^2}{8\mu}\equiv \delta\mu \ ,
\ee
where the choice of subscripts indicates that
$2\delta\mu_2$ is the
splitting between the Fermi surfaces for quarks 1 and 3 and
$2\delta\mu_3$ is that between the Fermi surfaces for quarks 1 and 2,
identifying $u,d,s$ with $1,2,3$.  (The prefactor $2$ in the equations defining
the $\delta\mu$'s is chosen to agree with the notation used in the analysis
of crystalline color superconductivity in a two flavor model~\cite{Alford:2000ze}, in which the
two Fermi surfaces were denoted by $\mu\pm\delta\mu$ meaning that they were
separated by $2\delta\mu$.)

As described in Refs. \cite{Rajagopal:2000ff,Alford:2002kj,Steiner:2002gx,Alford:2003fq},
BCS pairing introduces qualitative changes into the analysis of neutrality.  For example,
in the CFL phase $\mu_e=0$ and $\mu_8$ is nonzero and of order $M_s^2/\mu$.
This arises because the construction of a phase in which BCS pairing occurs between
fermions whose Fermi surface would be split in the absence of pairing can be
described as follows. First, adjust the Fermi surfaces of those fermions that pair to make
them equal. This costs a free energy price of order $\delta\mu^2\mu^2$.  And, it
changes the relation between the chemical potentials and the particle numbers,
meaning that the $\mu$'s required for neutrality can change qualitatively as
happens in the CFL example. Second, pair.
This yields a free energy benefit of order $\Delta_0^2\mu^2$, where $\Delta_0$
is the gap parameter describing the BCS pairing.
Hence, BCS pairing will only
occur if the attraction between  the fermions is large enough that
$\Delta_0 \gtrsim \delta\mu$.  In the CFL context, in which $\langle ud \rangle$,
$\langle us \rangle$ and $\langle ds \rangle$ pairing is fighting against the
splitting between the $d$, $u$ and $s$ Fermi surfaces described above, it
turns out that CFL pairing can occur if
$\Delta_0>4\delta\mu=M_s^2/(2\mu)$~\cite{Alford:2003fq},
a criterion that is
reduced somewhat by kaon condensation which
acts to stabilize CFL pairing~\cite{Kryjevski:2004jw}.
In this paper we are considering quark matter at densities that are
low enough ($\mu<M_s^2/(2\Delta_0)$) that CFL pairing is not possible.
The gap parameter $\Delta_0$ that would characterize the CFL
phase if $M_s^2$ and $\delta\mu$ were zero
is nevertheless an important scale in our problem, as it
quantifies the strength of the attraction between quarks.
Estimates of the magnitude of $\Delta_0$ are typically in the tens of MeV,
perhaps as large as 100 MeV~\cite{reviews}.  We shall treat $\Delta_0$ as a parameter,
and quote results for $\Delta_0=25$~MeV.

Crystalline color superconductivity can be thought of as the answer
to the question: ``Is there a way to pair quarks at differing Fermi
surfaces without first equalizing their Fermi momenta, given that
doing so exacts a cost?" The answer is ``Yes, but it requires Cooper
pairs with nonzero total momentum."  Ordinary BCS pairing pairs
quarks with momenta ${\bf p}$ and $-{\bf p}$, meaning that if the
Fermi surfaces are split at most one member of a pair can be at its
Fermi surface.  In the crystalline color superconducting phase,
pairs with total momentum $2\bf q$ condense, meaning that one member
of the pair has momentum ${\bf p}+{\bf q}$ and the other has
momentum $-{\bf p}+{\bf q}$ for some $\bf
p$~\cite{LOFF,Alford:2000ze}.  Suppose for a moment that only $u$
and $d$ quarks pair, making the analyses of a two-flavor model found
in
Refs.~\cite{Alford:2000ze,Bowers:2001ip,Casalbuoni:2001gt,Leibovich:2001xr,Kundu:2001tt,Bowers:2002xr,Casalbuoni:2003wh,Casalbuoni:2004wm}
(and really going back to Ref.~\cite{LOFF}) valid. We sketch the
results of this analysis in the following two paragraphs.

The simplest ``crystalline'' phase
is one in which only pairs with a single $\bf q$ condense, yielding a condensate
\be
\langle \psi_u(x) C \gamma_5 \psi_d(x) \rangle \propto \Delta \exp(-2i {\bf q}\cdot {\bf x})
\label{singleplanewave}
\ee
that is modulated in space like a plane wave.   Assuming that $\Delta\ll \delta\mu \ll \mu$,
the energetically favored value of $|{\bf q}|\equiv q$ turns out to be $q=\eta \delta\mu$, where
the proportionality constant $\eta$ is given by $\eta=1.1997$~\cite{LOFF,Alford:2000ze}.
If $\eta$ were 1, then the only choice of $\bf p$ for which a Cooper pair
with momenta $(-\bf p+\bf q,\bf p+\bf q)$ would describe two quarks each on
their respective Fermi surfaces would correspond to a quark on
the north pole of one Fermi surface and a quark on the south pole of the other.
Instead, with $\eta>1$, the quarks on each Fermi surface that can pair lie
on one ring on each Fermi surface, the rings having opening angle
$\psi_0=2\cos^{-1}(1/\eta)=67.1^\circ$.  The energetic calculation that determines $\eta$
can be thought of as balancing the gain in pairing energy as $\eta$ is increased
beyond $1$, allowing quarks on larger rings to pair, against the kinetic energy cost
of Cooper pairs with greater total momentum.   If the $\Delta/\delta\mu\rightarrow 0$
Ginzburg-Landau limit is not assumed, the pairing rings change from circular lines
on the Fermi surfaces into
ribbons of thickness $\sim\Delta$ and angular extent $\sim \Delta/\delta\mu$.
The condensate (\ref{singleplanewave}) carries a current, which is balanced by
a counter-propagating current carried by the unpaired quarks near
their Fermi surfaces that are not in the pairing ribbons. Hence, the state carries no net current.

After solving a gap equation for $\Delta$
and then evaluating the free energy of the phase with condensate (\ref{singleplanewave}),
one finds that this simplest ``crystalline'' phase is favored over two-flavor quark matter
with  either no pairing or BCS pairing only within  a narrow window
\be
0.707 < \frac{\delta\mu}{\Delta_0} < 0.754\ .
\label{LOFFwindow}
\ee
At the upper boundary of this window, $\Delta\rightarrow 0$ and one finds a second
order phase transition between the crystalline and unpaired phases.  At the lower boundary,
there is a first order transition between the crystalline and BCS paired phases.
The crystalline phase persists
in the weak coupling
limit only if $\delta\mu/\Delta_0$ is held fixed, within
the window (\ref{LOFFwindow}), while the standard weak-coupling limit
$\Delta_0/\mu\rightarrow 0$ is taken.  Looking  ahead to our context,
and recalling that in three-flavor quark matter $\delta\mu=M_s^2/(8\mu)$,
we see that at high densities one finds
the CFL phase (which is the three-flavor quark matter BCS phase) and
in some window of lower densities one finds a crystalline phase.
In the vicinity of the second order transition, where $\Delta\rightarrow 0$
and in particular where $\Delta/\delta\mu\rightarrow 0$ and, consequently given (\ref{LOFFwindow}),
$\Delta/\Delta_0\rightarrow 0$ a Ginzburg-Landau expansion
of the free energy order by order in powers of $\Delta$ is controlled.
Analysis within an NJL model shows that the results for $\Delta(\delta\mu)$  become
accurate in the limit $\delta\mu\rightarrow 0.754 \Delta_0$ where $\Delta\rightarrow 0$,
as must be the case, and show that the Ginzburg-Landau approximation underestimates
$\Delta(\delta\mu)$ at all $\delta\mu$~\cite{Alford:2000ze,Bowers:2002xr}.

The Ginzburg-Landau
analysis can then be applied to more complicated crystal structures in which
Cooper pairs with several different $\bf q$'s, all with the same length but pointing
in different directions, arise~\cite{Bowers:2002xr}.  This analysis indicates that a face-centered cubic
structure constructed as the sum of eight plane waves with $\bf q$'s pointing at
the corners of a cube is favored, but it does not permit a quantitative evaluation of
$\Delta(\delta\mu)$ in this case because it predicts a strong first order phase transition
between the crystalline and unpaired phase, at some $\delta\mu$ significantly larger
than $0.754 \Delta_0$, meaning that there is no value of $\delta\mu$ at which
the Ginzburg-Landau approximation is under control~\cite{Bowers:2002xr}.

Our purpose in this paper is to analyze a three-flavor analogue of
the simplest ``crystalline'' condensate (\ref{singleplanewave}), with and without
making the Ginzburg-Landau approximation $\Delta\ll\Delta_0$. We shall assume
weak coupling, namely $\Delta_0\ll \mu$, throughout.

The analysis of neutrality in three-flavor quark matter in a crystalline
color superconducting phase is very simple
in the Ginzburg-Landau limit in which $\Delta\ll \delta\mu$: because the construction of
this phase does {\it not} involve rearranging any  Fermi momenta prior to pairing,
and because the assumption $\Delta\ll \delta\mu$ implies that the pairing does not
significantly change any number densities, neutrality is achieved with the
same chemical potentials
$\mu_e=M_s^2/(4\mu)$ and $\mu_3=\mu_8=0$ as in unpaired quark matter, and
with Fermi momenta given in Eqs.~(\ref{pF1}), (\ref{pF2}), and
(\ref{pF3}) as in unpaired quark matter.
This result is correct only in the Ginzburg-Landau limit, but we expect that
deviations from it in our NJL analysis are small, and so we shall fix the chemical
potentials as in unpaired quark matter throughout this paper.

We consider a  condensate of the form,
\begin{equation}
\langle\psi_{\alpha i}(x)C \gamma_5 \psi_{\beta j}(x) \rangle \propto \sum_{I=1}^3 \Delta_I\,
e^{-2i{\bf q_I\cdot r}}\epsilon_{I\alpha\beta}\epsilon _{I i
j}\ , \label{condensate}
\end{equation}
where $\bf q_1$,  $\bf q_2$ and $\bf q_3$  and $\Delta_1$, $\Delta_2$  and
$\Delta_3$ are the wave vectors and gap parameters describing pairing between
the $(d,s)$, $(u,s)$ and $(u,d)$ quarks respectively, whose Fermi momenta are
split by $2\delta\mu_1$, $2\delta\mu_2$ and $2\delta\mu_3$ respectively.
From (\ref{pF3}), we see that $\delta\mu_2=\delta\mu_3=\delta\mu_1/2=M_s^2/(8\mu)$.
The condensate (\ref{condensate}), first analyzed in Ref.~\cite{Casalbuoni:2005zp}, is
the natural generalization of the condensate (\ref{singleplanewave}) to the
three-flavor setting, and the natural generalization of the CFL condensate (obtained
by setting $\bf q_1=\bf q_2=\bf q_3=0$) to the crystalline color superconductor
setting.

We shall make the further simplifying assumption
that $\Delta_1=0$.  Given that $\delta\mu_1$ is twice $\delta\mu_2$ or $\delta\mu_3$,
it seems reasonable that $\Delta_1\ll \Delta_2,\Delta_3$.   We leave a quantitative
investigation of the size of $\Delta_1$ to future work, however.  With $\Delta_1$
set to zero, the symmetry of the problem is such that we expect and find solutions with
$\Delta_2=\Delta_3\equiv\Delta$.  It should be noted, however, that these simplifications
are rigorous only in the Ginzburg-Landau limit in which $\Delta\ll \delta\mu$.  A full
investigation of cases in which $\Delta\sim \delta\mu$ requires investigating
the Ginzburg-Landau results  $\mu_3=\mu_8=0$, $\mu_e=M_s^2/(4 \mu)$,
$\Delta_1=0$, and $\Delta_2=\Delta_3=\Delta$ anew, as beyond the Ginzburg-Landau
approximation these are all assumptions, not results.

In an NJL model in which the quarks interact via a point-like four fermion interaction,
analyzed at the mean field level, if we assume that the only condensate is that
in Eq.~(\ref{condensate}) with $\Delta_1=0$
the interaction term added to the Lagrangian (\ref{lagr1})
can be written simply as
\begin{equation}
{\cal L}_{\Delta} = \sum_{I=2}^{3}\, \Delta_I^*\, e^{i 2 {\bf q_I \cdot
r }}\,\epsilon_{I\alpha\beta }\,\epsilon_{Iij}
\psi_{i\alpha}\,C\,\gamma_5\,\psi_{\beta j}~+~h.c.\; .\label{gapD0}
\end{equation}
Note that since the direction of one of the two wave vectors is
arbitrary, the quantities that have to be determined by minimizing
the free-energy are $\Delta$, the magnitude of the two wave vectors,
and the angle $\phi$ between  $\bf \hat q_2$ and $\bf \hat q_3 $.  We shall
see in the next section that, as in the two-flavor model, the magnitude of the wave vectors
is given in the Ginzburg-Landau approximation
by $|\bf q_2|=|\bf q_3|=\eta \delta\mu$ with $\eta=1.1997$.   And, we
shall see that, again as in the two-flavor model, $\Delta\rightarrow 0$
at $\delta\mu\rightarrow 0.754 \Delta_0$, corresponding to $M_s^2/\mu = 6.032\Delta_0$.
In the vicinity of this second order phase transition, the Ginzburg-Landau approximation
is controlled, as in the two-flavor model.
However, the angle $\phi$ between the wave vectors is a degree of freedom present here that has
no analogue in the two-flavor model
with condensate (\ref{singleplanewave}).
We shall determine it within the Ginzburg-Landau approximation
in the next section, and check this result in Section IV.

\section{Ginzburg-Landau analysis \label{GLsection}}

The free energy of the crystalline color superconducting phase can only
depend on the gaps $\Delta_I$ through $|\Delta_I |^2$, because the Lagrangian
has $U(1)$ symmetries associated with $u$-, $d$- and $s$-quark number conservation.
Of course, the condensate (\ref{condensate}) breaks these symmetries, but
it does so spontaneously.  Hence, when the free energy for the three-flavor
crystalline color superconductor is expanded in powers of the $\Delta_I$'s,
the only terms that can arise up to fourth order take the form $|\Delta_I |^2$,
$|\Delta_I |^4$, and $|\Delta_I|^2|\Delta_J|^2$ for $I\neq J$.  As we are setting
$\Delta_1=0$, this means that the Ginzburg-Landau expansion of the free energy
can be written
\begin{equation}
\label{omega} \Omega = \Omega_n +
\frac{\alpha}{2}(\Delta_2^2+\Delta_3^2) +
\frac{\beta}{4}(\Delta_2^4+\Delta_3^4)
  + \frac{\beta_{23}}{2}\Delta_2^2\Delta_3^2 +{\cal O}(\Delta^6)\;,
\end{equation}
where we have dropped the absolute value signs
as henceforth we will assume that the $U(1)$ symmetries are
broken such as to give real $\Delta$'s.  The contribution $\Omega_n$ is
the free energy for unpaired (``normal'') neutral quark matter with
$\mu_3=\mu_8=0$ and $\mu_e=M_s^2/(4\mu)$. Recall that
$\Delta_2$ describes up-strange pairs with momentum $2\bf q_2$, $\Delta_3$ describes
up-down pairs with momentum $2\bf q_3$,
and the three Fermi surfaces are ordered $d$
then $u$ then $s$ from biggest to smallest, each separated from
the next by $2\delta\mu = M_s^2/(4\mu)$.  The Ginzburg-Landau expansion
is controlled in the limit in which $\Delta_2/\delta\mu\rightarrow 0$ and
$\Delta_3/\delta\mu\rightarrow 0$.  In the present context, it was first analyzed
in Ref.~\cite{Casalbuoni:2005zp}.
The coefficients in the Ginzburg-Landau expansion (\ref{omega})
are given by the expressions
\bea \alpha &=& \frac{4\mu^2}{\pi^2}\left(
  -1 + \frac{\delta\mu}{2q}\ln\left|\frac{\delta\mu+q}{\delta\mu-q}\right|
  +\frac{1}{2}\ln\left|\frac{4(\delta\mu^2-q^2)}{\Delta_0^2}\right|
                    \right)\,,\nonumber\\
\beta &=& \frac{\mu^2}{2\pi^2}\Re e \!\int
\frac{d\hat{\bf p}}{4\pi}\frac{-1}{(i\epsilon-\hat{\bf p}\cdot{\bf q}+\delta\mu)^2
}
   =\frac{\mu^2}{\pi^2}\frac{1}{q^2-\delta\mu^2}\,,\nonumber\\
\beta_{23} &=& \frac{\mu^2}{2\pi^2}\Re e \!\int
\frac{d\hat{\bf p}}{4\pi}\frac{-1}{(i\epsilon-\hat{\bf p}\cdot{\bf q}_2-\delta\mu)
   (i\epsilon-\hat{\bf p}\cdot{\bf q}_3+\delta\mu)}\ ,\nonumber\\
 \quad & &  \label{beta23}
\eea
where $q\equiv|\bf q|$.
Note that in (\ref{omega}) the coefficients $\alpha$ and $\beta$ multiply
terms involving only $\Delta_2$ or only $\Delta_3$.  Since $\Delta_2$
and $\Delta_3$ separately each describe pairing between two fermions
with Fermi momenta separated by $2\delta\mu$, this means that
the expressions for $\alpha$ and $\beta$
are exactly as derived in the two-flavor analysis of Ref.~\cite{Bowers:2002xr}.
The coefficient $\beta_{23}$, derived in Ref.~\cite{Casalbuoni:2005zp}, is
instead characteristic of the 3-flavor case and represents the
interaction term between the two different condensates.   This is the only
term in which a dependence of the free energy on the angle $\phi$
between $\hat{\bf q}_2$ and $\hat{\bf q}_3$ can arise.
Note that we have
written explicitly the $i\epsilon$ prescription needed to evaluate the integral
expressions for both $\beta$ and $\beta_{23}$, which we have determined by
following carefully the derivation of (\ref{omega}) from the
Ginzburg-Landau expansion of the Green's function in the mean field
approximation~\cite{Bowers:2002xr}.
The $\beta_{23}$-integrand in Eq.~(\ref{beta23}) has singular points, making locating
its poles correctly in the complex plane crucial for the
evaluation of $\beta_{23}$.

To quadratic order, the free energy (\ref{omega}) includes no interaction
between the condensates $\Delta_2$ and $\Delta_3$, and so
is independent of $\phi$.
The expression for
$\alpha$ can be analyzed as in Ref.~\cite{Bowers:2002xr}.
As long as $\beta$ and $\beta_{23}$ are
positive, as we shall see below is the case,
there is a second order phase transition from unpaired
quark matter to a phase with nonzero $\Delta_2$ and $\Delta_3$
at the largest value of $\delta\mu$ for which $\alpha=0$ for some
value of $q$.  For larger $\delta\mu$, $\alpha>0$ for all $q$ and
the unpaired phase is stable.
The transition occurs at $\delta\mu=0.754\Delta_0$,
where modes with $q=\eta\delta\mu$ for $\eta=1.1997$ become unstable
to condensation, and $\Delta_2$ and $\Delta_3$ become nonzero.
At lower values of $\delta\mu$, modes in a band of $q$ have
$\alpha<0$ making them unstable, but
the mode with $q=\eta\delta\mu$ is
the one with the most negative $\alpha$ and
we therefore assume that the condensate involves only the modes
with $q_2=q_3=\eta\delta\mu$.

\begin{figure}[t]
\includegraphics[width=3.2in,angle=0]{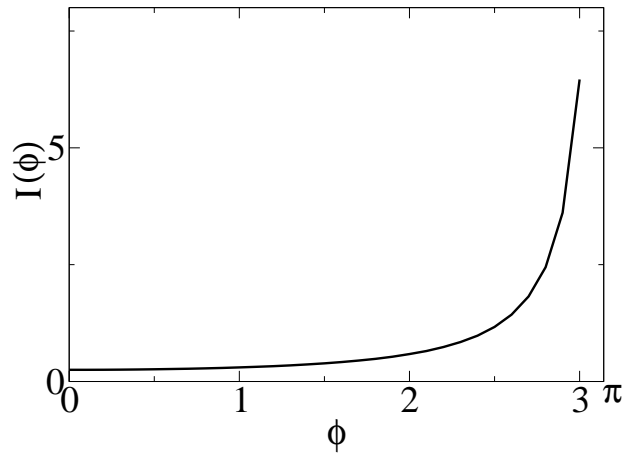}
\caption{ $I(\phi)$, defined in Eq.~(\ref{Iresult}), versus the
angle $\phi$ between the wave vectors ${\bf q}_2$ and ${\bf q}_3$. }
\label{funcbeta23}
\end{figure}

To evaluate the magnitude of $\Delta_2$ and $\Delta_3$ below
the second order phase transition where they become nonzero,
we need the quartic coefficients $\beta$ and $\beta_{23}$.
We see that although the location of the second order phase transition
is independent of $\phi$, the magnitude of the condensates
below the transition can depend on $\phi$ through $\beta_{23}$.
It is convenient to factor out the dependence of $\beta$ and
$\beta_{23}$ on $\delta\mu$ and $\mu$. For $q=1.1997\delta\mu$, then,
\be
\beta=2.276\frac{\mu^2}{\pi^2\delta\mu^2}
\label{betaresult}
\ee
and, similarly, we write
\be
\beta_{23}=\frac{2 \mu^2}{\pi^2\delta\mu^2}I(\hat{\bf q}_2,\hat{\bf q}_3)\ ,
\label{beta23result}
\ee
defining
\bea
I(\hat{\bf q}_2,\hat{\bf q}_3) &=&
  \Re e \int\frac{d\hat{p}}{4\pi}\frac{-1}{(i\epsilon-\eta\hat{\bf p}.\hat{\bf q}_2-1)
   (i\epsilon-\eta\hat{\bf p}.\hat{\bf q}_3+1)}\ .\nonumber\\
 \quad & &
  \label{Iresult}
\ee
From rotational invariance,
$I(\hat{\bf q}_2,\hat{\bf q}_3)$ depends only on $\phi$, the angle between
$\hat{\bf q}_2$ and $\hat{\bf q}_3$. $I(\phi)$ can
be evaluated numerically  and the result is plotted in
Fig.\ref{funcbeta23}. We note that $I(\phi)$ is an increasing
function of $\phi$ and diverges at $\phi=\pi$, i.e. when the two $\bf q$
vectors are antiparallel.  The minimum value of $I$ occurs at $\phi=0$,
for parallel $\bf q$ vectors.

\begin{figure*}[t]
\leavevmode
\begin{center}
\epsfxsize=17cm \epsfbox{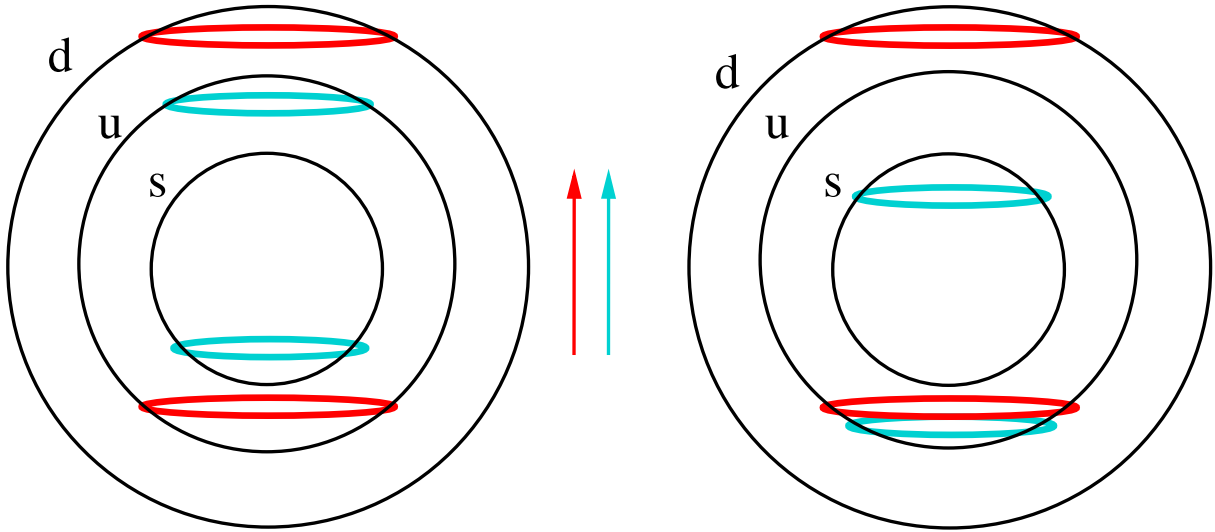}
\end{center}
\caption{Sketch showing where on the Fermi surfaces pairing occurs for condensates
in which $\bf q_2$ and $\bf q_3$ are parallel (left panel) or antiparallel (right panel).  The
dark grey (red online) rings on the $d$ and $u$ Fermi surfaces indicate those
quarks that contribute the most to the $\langle ud \rangle$ condensate
with gap parameter $\Delta_3$ and wave vector $\bf q_3$, which points upward in both panels.
The light grey (teal online) rings on the $u$ and $s$ Fermi surfaces indicate
those quarks that contribute most to the $\langle us \rangle$ condensate
with gap parameter $\Delta_2$ and wave vector $\bf q_2$, which points upward in the
left panel and downward in the right panel.  For
illustrative purposes, we have greatly exaggerated the splitting between the Fermi surfaces,
$\delta\mu/\mu$ relative to the values that we shall employ in the calculations
reported in Section V.}
\label{ribbonsfigure}
\end{figure*}

We now look for minima of the free energy with
$\Delta_2=\Delta_3=\Delta$, where the free energy can then be
written as
\begin{equation}
\Omega=\Omega_n+\alpha\Delta^2+\frac{1}{2}(\beta+\beta_{23})\Delta^4
\;.
\end{equation}
The expression for $\alpha$ in Eq.~(\ref{beta23})  yields the result
that for $q=\eta\delta\mu$, $\alpha<0$ for $\delta\mu <
0.7544\Delta_0$, corresponding to $M_s^2/\mu < 6.035 \Delta_0 \simeq
151$~MeV, where we have taken $\Delta_0=25$~MeV.  For values of
$\delta\mu$ where $\alpha$ is negative, the solution is found at \be
\Delta^2=\frac{|\alpha|}{\beta+\beta_{23}} \label{DeltaGL} \ee with
\be \Omega=\Omega_n - \frac{\alpha^2}{2(\beta + \beta_{23})}\ .
\label{OmegaGL} \ee These expressions can be evaluated using
$\alpha$, $\beta$ and $\beta_{23}$  from Eqs.~(\ref{beta23}),
(\ref{betaresult}) and (\ref{beta23result}) respectively. Since
$\beta+\beta_{23}$ is positive for all choices of the angle $\phi$,
we see that the transition from the unpaired state to the
crystalline phase with the simple ``crystal'' structure that we have
employed is second order for all choices of $\phi$. The best choice
of $\phi$ is the one that minimizes $\Omega$, meaning the choice
with the smallest $\beta_{23}$.  From Fig.~\ref{funcbeta23} we see
that this corresponds to $\phi=0$, with ${\bf q}_2 \parallel {\bf
q}_3$. We shall provide plots of $\Delta$ and $\Omega$ versus
$M_s^2/\mu$ for various values of $\phi$ in Section V, where we
shall compare these results to results obtained without making the
Ginzburg-Landau approximation.

The free energy (\ref{omega}) can also be used to analyze the free energy
of a two-flavor crystalline  phase with a single-plane wave ``crystal structure''
in three flavor quark matter.  Setting $\Delta_2=0$ and $\Delta_3=\Delta$ (or
equivalently setting $\Delta_3=0$ and $\Delta_2=\Delta$) we find a solution
with $\Delta^2=|\alpha|/\beta$ and $\Omega-\Omega_n=-\alpha^2/(4\beta)$.
Like the solution with $\Delta_2=\Delta_3$, this solution is
neutral in the Ginzburg-Landau limit.  The solution with $\Delta_2=\Delta_3$
has a lower free energy than that with $\Delta_2=0$ if $(\beta+\beta_{23})<2\beta$,
meaning $\beta_{23}<\beta$.
One can generalize to  other solutions, and rule them out, by
writing $(\Delta_2,\Delta_3)$
as $(\Delta_r \sin\theta,\Delta_r\cos\theta)$ and then rewriting Eq.~(\ref{omega})
as
\be
\Omega=\Omega_n + \frac{\alpha}{2}\Delta_r^2 + \frac{\beta}{4}\Delta_r^4
+ \left(\frac{\beta_{23}-\beta}{8}\right)\Delta_r^2 \sin^2(2\theta)\ .
\ee
As long as $\alpha<0$, there is a minimum with $\theta=\pi/4$ (namely
$\Delta_2=\Delta_3$) if $\beta_{23}<\beta$ and minima
at $\theta=0$ and $\theta=\pi/2$ if $\beta_{23}>\beta$.
From Eqs.~(\ref{betaresult}) and (\ref{beta23result}) we see that $\beta_{23}<\beta$
if $I(\phi)<1.138$, and from Fig.~\ref{funcbeta23} we see that this occurs for
$\phi<2.485$ radians, or $\phi<142.4^\circ$.

Although the divergence of $\beta_{23}$ at $\phi=\pi$ is not physically relevant,
since for $\beta_{23}>\beta$ the lowest free energy solution has $\Delta_2$
or $\Delta_3$ vanishing, it is still worth gaining a qualitative understanding
of the origin of the divergence.  We  see in Fig.~\ref{ribbonsfigure} that
there are two pairing rings on the up quark Fermi surface, because some
up quarks pair with down quarks forming Cooper pairs with wave vector $2\bf q_3$
and other up quarks pair with strange quarks
forming Cooper pairs with wave vector $2\bf q_2$.
However, as shown in the
right panel of Fig.~\ref{ribbonsfigure}, if  $\phi=\pi$ the two pairing rings on the
up quark Fermi surface are close to coincident.  In the weak-coupling limit
in which $\delta\mu/\mu\rightarrow 0$ (and $\Delta_0\rightarrow 0$
with $\delta\mu/\Delta_0$ fixed) these two rings become precisely
coincident.  We attribute the divergence in $\beta_{23}$,
which indicates that antiparallel wave vectors pay
an infinite free energy price and hence are forbidden, to the coincidence
of these two pairing rings. Loosely speaking, it  is as if these up quarks do not know whether to
pair with their putative strange or down partners and so do neither.
In contrast, if $\phi=0$ as in the left panel of the Figure,
the two pairing rings on the up Fermi surface are as far apart as they can be,
and $\beta_{23}$ and the free energy of the state are minimized.
This qualitative understanding also highlights that it is  only in the
strict Ginzburg-Landau and weak coupling limits that the
cost of choosing antiparallel wave vectors diverges.  If $\Delta/\delta\mu$ is small
but nonzero, the pairing regions are ribbons on the Fermi surfaces instead of lines.
And, if $\delta\mu/\mu$ is small but not taken to zero (as of course is the case in
Fig.~\ref{ribbonsfigure}) then the two ribbons on the up Fermi surface will
have slightly different diameter, as the Figure indicates. This means that we expect
that if we do a calculation at small but nonzero $\Delta_0\sim\delta\mu$, and do
not make a Ginzburg-Landau expansion, we should find some
free energetic penalty for choosing $\phi=\pi$, but not a divergent one.
We shall set up this calculation in Section IV and
see this expectation confirmed in Section V.

\section{NJL analysis without Ginzburg-Landau approximation \label{HDETsection}}

The two-flavor crystalline color superconducting phase with a single plane-wave
``crystal'' structure (\ref{singleplanewave}) has been analyzed in a variety of ways without making
a Ginzburg-Landau approximation, going back to the work of Fulde and Ferrell~\cite{LOFF}.
In the QCD context, it was analyzed using a variational method in Ref.~\cite{Alford:2000ze}, using
a diagrammatic method employing a modification of the Nambu-Gorkov formalism
in Refs.~\cite{Bowers:2001ip,Kundu:2001tt}, and using the Nambu-Gorkov formalism
simplified via the High Density Effective Theory in Ref.~\cite{Casalbuoni:2001gt}.

In the conventional Nambu-Gorkov formalism as applied to ordinary BCS pairing,
one defines an eight-component Nambu-Gorkov spinor
\be
\Psi({\bf p}) = \left(\begin{array}{l} \psi({\bf p}) \\ \bar\psi^T(-{\bf p})
\end{array}\right)\ ,
\ee
such that in this basis the pairing between fermions
with momentum ${\bf p}$ and ${\bf -p}$ is described by an off-diagonal term
in the fermion propagator.  The condensate (\ref{singleplanewave}), however,
describes pairing between  $u$ quarks with momentum ${\bf p}+{\bf q}$
and $d$ quarks with momentum $-{\bf p}+{\bf q}$.  This could
be described via a propagator with terms in it that are off-diagonal in {\it momentum} space,
rather than merely off-diagonal in ``Nambu-Gorkov space''. However, it is {\it much}
easier to change to a
basis in which the Nambu-Gorkov spinor is written as~\cite{Bowers:2001ip}
\be
\Psi({\bf p}) = \left(\begin{array}{l} \psi_u({\bf p}+{\bf q})\\ \psi_d({\bf p}-{\bf q}) \\
\bar\psi_u^T(-{\bf p}-{\bf q})\\ \bar\psi_d^T(-{\bf p}+{\bf q})
\end{array}\right)\ .
\label{NGspinor2a} \ee The pair condensate is then described by
terms in the fermion propagator that are off-diagonal in
Nambu-Gorkov space, occurring in the $\psi_u$-$\bar\psi^T_d$ and
$\psi_d$-$\bar\psi_u^T$ entries. In this basis the fermions that
pair have ${\bf p}$ and ${\bf -p}$, making the propagator diagonal
in ${\bf p}$-space and the calculation tractable. One  must always
keep in mind that it is ${\bf p}+{\bf q}$ and $-{\bf p}+{\bf q}$
that are the momenta of the fermions that pair, not ${\bf p}$ and
${\bf -p}$. The variable ${\bf p}$ is an integration variable: in
the gap equation or in the expression for the free energy,
integrating over ${\bf p}$ sums the contributions of all the
fermions although of course it turns out that only those lying near
ribbons on the Fermi surfaces contribute significantly. Since ${\bf
p}$ {\it is} an integration variable, we are free to change
variables, for example rewriting the Nambu-Gorkov spinor as \be
\Psi({\bf p}) = \left(\begin{array}{l} \psi_u({\bf p})\\ \psi_d({\bf p}-2{\bf q}) \\
\bar\psi_u^T(-{\bf p})\\\bar\psi_d^T(-{\bf p}+2{\bf q})
\end{array}\right)\ .
\label{NGspinor2b}
\ee

The form of the Nambu-Gorkov spinor (\ref{NGspinor2b})
 immediately suggests that we analyze our three-flavor crystalline phase
with condensate (\ref{condensate}) with $\Delta_1$ set to zero
by introducing the Nambu-Gorkov spinor
\be
\Psi({\bf p}) = \left(\begin{array}{l} \psi_u({\bf p})\\ \psi_d({\bf p}-2{\bf q}_3) \\
\psi_s({\bf p}-2{\bf q}_2)\\
\bar\psi_u^T(-{\bf p})\\ \bar\psi_d^T(-{\bf p}+2{\bf q}_3) \\
\bar\psi_s^T(-{\bf p}+2{\bf q}_2)
\end{array}\right)\ .
\label{NGspinor}
\ee
Furthermore, it also indicates that it will not be possible to use this
method of calculation if $\Delta_1$ were kept nonzero, except
for the special case in which ${\bf q}_1={\bf q}_2 - {\bf q}_3$. (That is, except in
this special case which is far from sufficiently generic, it will not be
possible to choose a Nambu-Gorkov basis such that one obtains
a propagator that is diagonal in some momentum variable ${\bf p}$.)
It is thus fortunate that, as
explained in Section II, it is reasonable on physical grounds to begin with
$\Delta_1=0$, as we do throughout this paper.  Finally, it seems unlikely that  this
method can be employed to analyze more complicated crystal structures
analogous to the face-centered  cubic structure that is favored in the
two-flavor case~\cite{Bowers:2002xr}. Investigating such possibilities
is feasible in the Ginzburg-Landau approximation, but the condensate
that we are analyzing, with $\Delta_1=0$ and $\Delta_2$ and $\Delta_3$
each multiplying a single plane wave, is the most complex example that we
currently know how to analyze without making the Ginzburg-Landau approximation.

We now implement the calculation in the basis (\ref{NGspinor}) using the High
Density Effective Theory formalism of Refs.~\cite{Casalbuoni:2001gt,Nardulli:2002ma},
valid in the weak-coupling limit in which $\Delta_0\ll\mu$. We Fourier decompose the
fermionic fields in the following nonstandard fashion:
\begin{equation} \psi_{i\alpha}(x)= e^{-i {\bf k_i \cdot x}}\int\frac{d{\bf
n}}{4\pi}e^{-i\mu{\bf n}\cdot{\bf x}}\,\left(\psi_{{i\alpha},\bf
n}(x)+\psi^-_{{i\alpha},\bf n}(x)\right)\label{decomp} \, ,
\end{equation}
where ${\bf n}$ is a unit three-vector whose direction is integrated
over, where ${\bf k_i}$ are three fixed vectors, one for each
flavor, that we shall specify momentarily and where
$\psi_{{i\alpha},\bf n}(x)$ (resp. $\psi^-_{{i\alpha},\bf n}(x)$)
are positive (resp. negative) energy projections of the fermionic
fields with flavor $i$ and color $\alpha$, as defined in
Refs.~\cite{Casalbuoni:2001gt,Nardulli:2002ma}. In the usual HDET
approximation~\cite{Nardulli:2002ma}, the vectors ${\bf k}_i$ are
zero and the field $\psi_{{i\alpha},\bf n}(x)$ is used to describe
quarks in a patch in momentum space in the vicinity of  momentum
${\bf p}=\mu{\bf n}$. The introduction of the shift vectors  means
that now $\psi_{{i\alpha},\bf n}(x)$ describes quarks with momenta
in a patch in the vicinity of momentum $\mu{\bf n} +{\bf k}_i$, with
${\bf k}_i$ different for different flavors. To reproduce
(\ref{NGspinor}), then, it appears that we should choose \bea
{\bf k}_u &=& 0 \nonumber\\
{\bf k}_d &=& 2 {\bf q}_3 \nonumber\\
{\bf k}_s &=& 2 {\bf q}_2 \ .
\label{choiceofk1}
\eea
We shall see this choice emerge in a different way below.

\begin{figure*}[t]
\leavevmode
\begin{center}
\epsfxsize=17cm \epsfbox{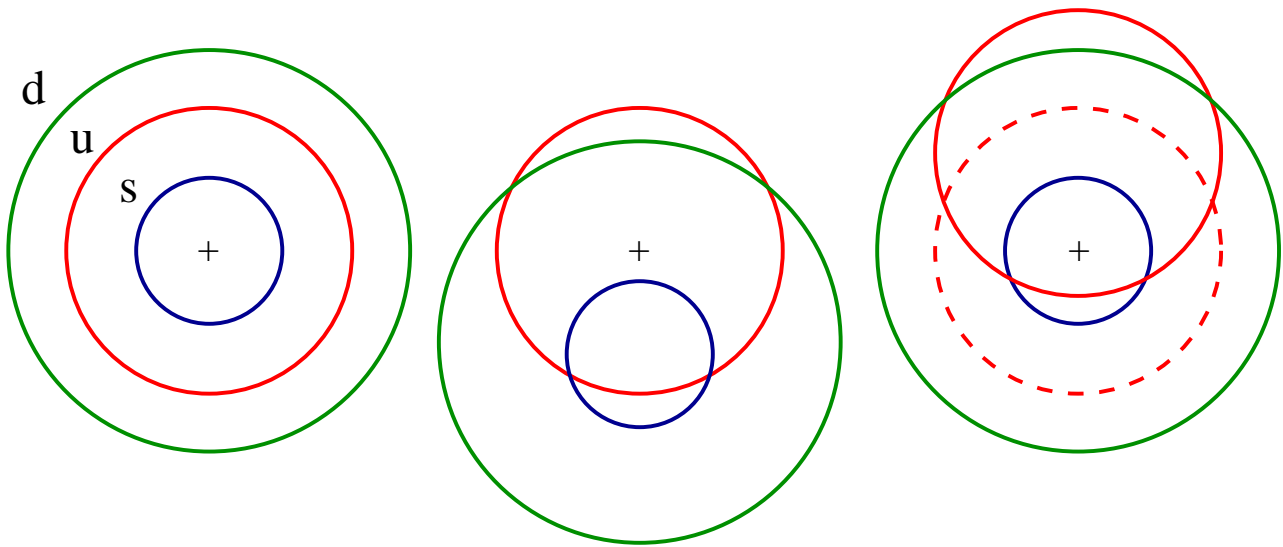}
\end{center}
\caption{Sketches showing how different choices of the shift vectors ${\bf k_i}$
can be interpreted.  The  left panel shows unshifted $d$, $u$ and $s$ Fermi surfaces,
in the absence of pairing.
As in Fig.~\ref{ribbonsfigure},
we have exaggerated the magnitude of $\delta\mu/\mu$ for illustrative purposes.
Now, suppose we have a condensate with
${\bf q}_2={\bf q_3}\equiv {\bf q}$, with ${\bf q}$ a vector pointing upwards.
In order to describe the
$(\psi_{{\bf n},d},\psi_{-{\bf n},u})$ and $(\psi_{{\bf n},s},\psi_{-{\bf n},u})$ pairing,
we shift the $u$, $d$ and $s$ Fermi surfaces by $-{\bf k}_u$, $-{\bf k}_d$ and $-{\bf k}_s$
respectively (since $\psi_{{\bf n},i}$ describes quarks in the vicinity of $\mu{\bf n}+{\bf k}_i$)
and then reflect (i.e. take ${\bf n} \rightarrow -{\bf n}$) the $u$ Fermi surface.
The middle panel shows the outcome if we follow this procedure with
shift vectors given by (\ref{choiceofk1}).
The
$u$ Fermi surface is left unshifted (meaning its inversion
is invisible in the Figure), and the $d$ and $s$ Fermi surfaces are shifted downwards
by $2{\bf q}$.   In the right panel, we instead
choose ${\bf k}_u=2{\bf q}$, which according to (\ref{choiceofk2}) then
requires ${\bf k}_d={\bf k}_s=0$.  The $d$ and $s$ Fermi surfaces are
unshifted.
The $u$ Fermi surface has been shifted downwards by $2{\bf q}$ and
then inverted, making it look as if it was shifted upwards.
The location
of the rings on the Fermi surfaces where pairing occurs are determined by the places
where the circles cross.  In both middle and right panels,
the pairing rings on the $d$ and $s$ Fermi surface occur at their intersections with
the $u$  Fermi surfaces whereas the pairing rings on the $u$ Fermi
surface are antipodal to where these intersections appear in the Figure, since the
$u$ Fermi surface was inverted in constructing the Figure.  Thus, both the right and middle panels
of this figure correspond to the pairing sketched in the left panel of Fig.~\ref{ribbonsfigure}.
The same calculation can be done by integrating over the momentum
variable of either the middle or right panel; the difference between panels
is just a change of integration variable.
The origins of the momentum variables are indicated by the $+$ in each panel.
Note that if we had instead chosen to describe the
$(\psi_{{\bf n},u},\psi_{-{\bf n},d})$ and $(\psi_{{\bf n},u},\psi_{-{\bf n},s})$ pairing,
both the middle and right panels would look inverted relative to those given but
this difference also
corresponds to a change of integration variable, in this case ${\bf n} \leftrightarrow -{\bf n}$.
}
\label{fermiloff}
\end{figure*}

\begin{figure*}[t]
\leavevmode
\begin{center}
\epsfxsize=18cm \epsfbox{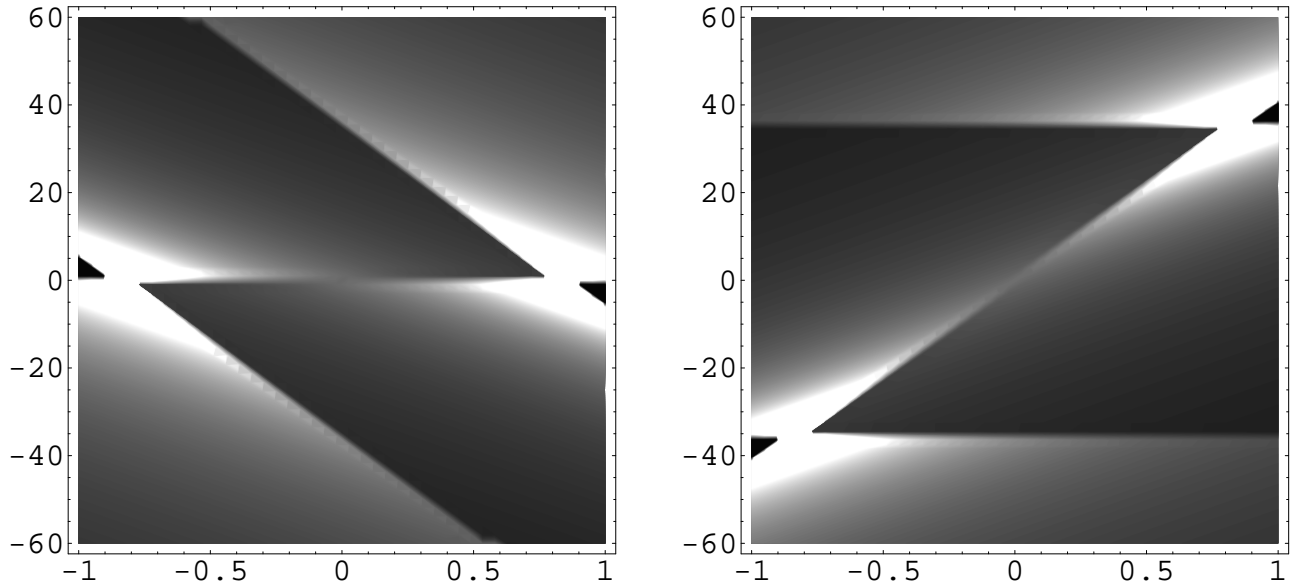}
\end{center}
\caption{Gap equation integrand $f$, defined in Eq.~(\ref{fdefn}), as a function of
momentum.  In each panel, the horizontal axis is $\cos\theta$ where $\theta$
is the polar angle specified by $\bf n$, the vertical axis is $\ell_\parallel$ in MeV,
and the grey scale indicates the value of $f$: black corresponds to $f=0$ and white
to the largest values of $f$.  In both panels, $\mu=500$~MeV and $\delta\mu=17.7$~MeV, which
is $0.707 \Delta_0$ for $\Delta_0=25$~MeV, the interaction strength that we shall use
in the next section. In both panels,
${\bf q}_2={\bf q}_3\equiv {\bf q}$ with $q=1.20\, \delta\mu = 21.2$~MeV and $\bf q$ pointing
in the $z$-direction. And, in both panels $\Delta=1$~MeV.  In the left panel, ${\bf k}_u=0$ and
${\bf k}_d={\bf k}_s=2{\bf q}$ as in the middle panel of Fig.~\ref{fermiloff}.
The unshifted $u$ Fermi surface is centered in momentum space, meaning
that it appears in the left panel as a horizontal line at $\ell_\parallel=0$.
The shifted $d$ and $s$ Fermi surfaces appear as diagonal lines, with the
shifted $d$ Fermi surface inside the $u$ Fermi surface at the north pole ($\cos\theta=1$)
and the shifted $s$ Fermi surface outside the $u$ Fermi surface at the south pole.
In the right panel, ${\bf k}_u=2{\bf q}$ and
${\bf k}_d={\bf k}_s=0$ as in the right panel of Fig.~\ref{fermiloff}. The unshifted
$d$ and $s$ Fermi surfaces are centered in momentum space, meaning that
they appear in the right panel as horizontal lines at $\ell_\parallel=\pm 2\delta\mu$.
(Note that $\ell_\parallel$ is measured relative to where the
unshifted $u$-Fermi surface would have been,
shown as a dashed circle in
the right panel of Fig.~\ref{fermiloff} that corresponds to $\ell_\parallel=0$
in the right panel here.)  The shifted $u$ Fermi surface appears in the right panel here
as a diagonal line, outside the $d$ Fermi surface at the north pole
and inside the $s$ Fermi surface at the south pole.
Pairing is most important in the bright white regions,
centered where the shifted Fermi surfaces cross.
The sketches provided in Fig.~\ref{ribbonsfigure} and
particularly in Fig.~\ref{fermiloff} serve to help
visualize the ``momentum-space geometry'' and pairing
regions depicted in the present figure.
}
\label{pairingregions}
\end{figure*}

Substituting the expression (\ref{decomp}) in Eqs.~(\ref{lagr1}) and
(\ref{gapD0}) and neglecting the contribution of antiparticles, the
full Lagrangian reads
\begin{equation}
\begin{split}
{\cal L}=& \int\frac{d{\bf n}}{4\pi}\, \Biggl[ \psi_{{\bf
n},i\alpha}^\dagger\left(i V \cdot \de^{\alpha\beta}_{ij} +
\delta\mu_{i}({\bf n})\delta^{\alpha\beta}\delta_{ij} \right)
\psi_{{\bf n},\beta j}\\
&+ \Bigl(\sum_{I=2}^{3}\,\Delta_I  e^{i 2 {\bf q_I \cdot r
}}\,\epsilon_{I\alpha\beta }\,\epsilon_{Iij} \psi_{{\bf
n},i\alpha}\,C\,\gamma_5\,\psi_{{-\bf n},\beta j} e^{-i  {\bf
({\bf k}_i+{\bf k}_j) \cdot {\bf r} }} \\ &+ h.c.\Bigr)\Biggr] \label{L11}\, ,
\end{split}
\end{equation}
where   $\delta\mu_{i}({\bf  n})= P^F_i-\mu - {\bf k_i \cdot n}$ and
where the four vectors $V^\nu$ and $\bar V^\nu$ (the latter used
only below) are defined by $V^\nu=(1,{\bf n})$ and $\bar V^\nu=(1,-{\bf n})$.
We now see that
we can get rid of the space dependence in the gap term by choosing the
shift vectors $\bf k_i $ so that they satisfy
\bea
\bf{k_u+k_d} &=& 2 {\bf q}_3   \nonumber \\
\bf{k_u+k_s} &=& 2{\bf q}_2\ . \label{choiceofk2}
\eea
Because the ${\bf k}$'s were introduced arbitrarily in the decomposition (\ref{decomp}),
the calculation could in principle be done with any choice of ${\bf k}$'s.  However, eliminating
the space dependence in the gap term is an enormous simplification, equivalent to
yielding a propagator that is diagonal in momentum space, and is what makes the
calculation tractable. So, we shall always choose ${\bf k}$'s satisfying (\ref{choiceofk2}).
According to (\ref{choiceofk2}), if we choose ${\bf k}_u=0$, we recover (\ref{choiceofk1}).
 However, ${\bf k}_u$ can be chosen arbitrarily as long as
${\bf k}_d$ and ${\bf k}_s$ are then chosen to satisfy (\ref{choiceofk2}).
This means that the choices of ${\bf k}$'s that get rid of the space dependence in
the gap term are given by any combination related  to (\ref{choiceofk1}) by adding
any vector to ${\bf k}_u$ and subtracting the same vector from ${\bf k}_d$ and ${\bf k}_s$.
The geometric interpretation
of two examples of choices of ${\bf k}$'s is described in Fig,~\ref{fermiloff}.
As this figure illustrates,
the freedom to shift ${\bf k}_u$ while keeping
${\bf k}_u+{\bf k}_d$ and ${\bf k}_u+{\bf k}_s$ fixed corresponds
to the freedom to shift integration variable, for example as we did in
going from (\ref{NGspinor2a}) to (\ref{NGspinor2b}).
In obtaining the results that we shall plot in Section V, we shall use
the choice (\ref{choiceofk1});
however, we have checked numerically that different
choices of ${\bf k}_u$ with
${\bf k}_d$ and ${\bf k}_s$ satisfying (\ref{choiceofk2}) yield the
same results for the gap parameter and free energy. As Fig.~\ref{pairingregions}
below indicates, these different choices yield quite different intermediate stages
to the calculation so the fact that we find the expected agreement between
them is a nontrivial check of our numerics.

We can now employ the  Nambu-Gorkov basis defined in detail in Ref.~\cite{Casalbuoni:2004tb}
given by
\begin{equation}
\chi_A=\frac{1}{\sqrt{2}}\left(\begin{array}{c}
  \psi_{\bf n} \\
  C\,\psi^*_{- \bf n}
\end{array}\right)_A \label{chi}\ ,
\end{equation}
where $A=1\ldots 9$ is a color-flavor index running over the nine quarks (three
colors; three flavors) and where the $\psi_{\bf n}$ fields are defined via (\ref{decomp}) with
shift vectors chosen as in Eq.~(\ref{choiceofk1}).  In this basis, the full Lagrangian
can be written in the compact form
\begin{equation}
{\cal L}=\sum_{\bf n}\,\chi^\dagger_A~S^{-1}_{A,B}({\bf n})~\chi_B
\, ,\label{total}
\end{equation}
with
\begin{equation}
S^{-1}_{A,B}=\left(
\begin{array}{cc}
  \left( V\cdot \ell\, + \delta\mu_{A}({\bf  n}) \right) \delta_{AB} & -\Delta_{AB} \\
  -\Delta_{AB} &  (\bar V\cdot \ell\, - \delta\mu_{A}({\bf -n}))\delta_{AB}
\end{array}
\right)\, ,\label{inverse}
\end{equation}
where $\ell_\nu = (\ell_0,\ell_\parallel {\bf n}) $ is a four-vector. Here,
$\ell_\parallel$ is
the ``radial'' momentum component of $\psi$, parallel to ${\bf n}$.
In HDET, the  momentum of a fermion is written as $(\mu+\ell_\parallel){\bf n}$, with
the integration over momentum space separated into an angular integration
over ${\bf n}$ and a radial integration over $-\delta < \ell_\parallel < \delta$.
Here, the cutoff $\delta$ must be  smaller than $\mu$
but  must be much larger than $\Delta_0$, $\delta\mu$ and $\Delta$.  In the
results we plot in Section V, we shall take $\mu=500$~MeV, $\delta=300$~MeV
and $\Delta_0=25$~MeV.

{}From the Lagrangian  (\ref{total}), following a derivation
analogous to that in Ref.~\cite{Alford:2004hz},
the thermodynamic potential per unit volume
can be evaluated to be
 \bea
 \Omega &=&
- \frac{\mu^2}{4 \pi^2}\sum_{a=1,18}\int_{-\delta}^{+\delta} d
\ell_\parallel \int \frac{d{\bf n}}{4\pi} ~|E_a({\bf n},\ell_\parallel)|\nonumber\\
& &+\frac{2\Delta^2}{G}-
\frac{\mu_{e}^4}{12 \pi^2} \label{Omega}
\eea
where we have set $\Delta_2=\Delta_3=\Delta$ and
where   $G$ is a
coupling constant chosen in such a way that $\Delta_0=25$~MeV
in the CFL phase found
at $\mu=500$~MeV with $M_s=0$.
In this expression, the $E_a$  are the energies
of the quasiparticles in this phase. They are given by the 18 roots
of $\det S^{-1}=0$, seen as an equation for $\ell_0$, with $S^{-1}$ the
Nambu-Gorkov inverse propagator given in (\ref{inverse}).
The quasiparticle energies are functions of $\ell_{\|}$ and ${\bf n}$, and also depend
on the gap parameter $\Delta$ and the wave vectors $\bf k_i$.
The doubling of degrees of freedom in the Nambu-Gorkov formalism
means that the 18 roots come in pairs whose
energies are related by $E_a({\bf n},\ell_\parallel) = E_b(-{\bf n},\ell_\parallel)$.
One set of nine roots describes
$(\psi_{{\bf n},d},\psi_{-{\bf n},u})$ and $(\psi_{{\bf n},s},\psi_{-{\bf n},u})$ pairing,
while the other set describes
$(\psi_{{\bf n},u},\psi_{-{\bf n},d})$ and $(\psi_{{\bf n},u},\psi_{-{\bf n},s})$ pairing.
Since ${\bf n}$ is integrated over,
the free energy can be evaluated by doing the sum in (\ref{Omega})
over either set of nine roots, instead of over all 18, and multiplying the sum by two.

In order to determine the lowest free energy state, we need to minimize
the free energy $\Omega$ given in Eq.~(\ref{Omega}) with respect to
the gap parameter $\Delta$ and with respect to $\phi$, the angle between $\hat{\bf q}_2$
and $\hat{\bf q}_3$.   One could also simultaneously minimize with
respect to $\mu_e$, $\mu_3$ and $\mu_8$.  And, one
could allow $\Delta_2\neq \Delta_3$ and minimize
with respect to the two gap parameters separately. However, in the results
that we shall present in the next section we shall fix $\Delta_2=\Delta_3=\Delta$,
$\mu_e=M_s^2/(4\mu)$
and $\mu_3=\mu_8=0$, as is correct for small $\Delta$ and as we
have done in the Ginzburg-Landau analysis of Section III.

Before turning  to comparing results obtained
from the calculation presented in this section to
those obtained with the Ginzburg-Landau approximation
developed in Section III, we close this section by calculating explicitly how
the free energy $\Omega$ of Eq.~(\ref{Omega})
manifests the qualitative features
described in Section II and sketched in Fig.~\ref{ribbonsfigure}, with
pairing occurring along ribbons of the Fermi surfaces.
The easiest way to find the regions of momentum
space in which pairing is important is to analyze the gap equation, obtained
by varying $\Omega$ with respect to $\Delta$.  This takes the form
\be
\Delta \propto
\int_{-\delta}^{+\delta} d\ell_\parallel \int d{\bf n}\,
f({\bf n},\ell_\parallel)
\label{gapeqn}
\ee
with
\be
f({\bf n},\ell_\parallel)=\sum_{a=1}^{9} \frac{\partial E_a}{\partial \Delta} {\rm Sign}(E_a)\ .
\label{fdefn}
\ee
(Either set of nine quasiparticle energies could be chosen, but to make the
comparison with Fig.~\ref{fermiloff} we have used those describing
$(\psi_{{\bf n},d},\psi_{-{\bf n},u})$ and $(\psi_{{\bf n},s},\psi_{-{\bf n},u})$ pairing.)
In Fig.~\ref{pairingregions}, we plot $f$, the integrand in the gap equation, as a function
of $\ell_\parallel$ and $\cos \theta$ where $\theta$ is the polar angle specified
by ${\bf n}$.  (A plot of the integrand in the expression for $\Omega$ in Eq.~(\ref{Omega})
evaluated with $\Delta$ minus that evaluated with $\Delta=0$ yields a very similar
figure.)
We have plotted $f(\cos\theta,\ell_\parallel)$
for two different choices of the shift vectors ${\bf k}_i$, corresponding
to the middle and right panels in  Fig.~\ref{fermiloff}.  The differences
between the two panels of Fig.~\ref{pairingregions} come entirely from the
different choices of shift vectors; both panels correspond to the same
condensate,
with ${\bf q}_2 \parallel {\bf q_3}$ as in
the left panel of Fig.~\ref{ribbonsfigure}.
And, we find excellent agreement when we integrate $f$ depicted in
either the left or the right panel of Fig.~\ref{pairingregions} to obtain
the right-hand side of the gap equation (\ref{gapeqn}),  and similar agreement when we do the
integral in Eq.~(\ref{Omega})  needed to evaluate the free energy $\Omega$ with
either choice of shift vectors.

In both panels of Fig.~\ref{pairingregions}, the
bright white pairing regions near where the shifted and unshifted Fermi surfaces cross
are clearly visible, as are the jet black blocking regions near the north
and south poles of the Fermi surfaces at $\cos\theta=\pm 1$ where no pairing
occurs.  The pairing
regions are centered at $\theta=67.1^\circ/2$
and $\theta=180^\circ - 67.1^\circ /2$, corresponding to $\cos\theta=\pm 0.833$.
The
dark but not black regions between Fermi surfaces are places where either $u$-$d$ or
$u$-$s$ pairing is blocked, but the other is allowed.  Note that even though
the formal pairing regions (regions where $f\neq 0$) extend far from the Fermi surfaces, the
bright white
regions where the maximal value of $f$ is attained are localized near where Fermi
surfaces cross.  So far, all as expected.

\begin{figure*}[t]
\includegraphics[width=2.7in,angle=-90]{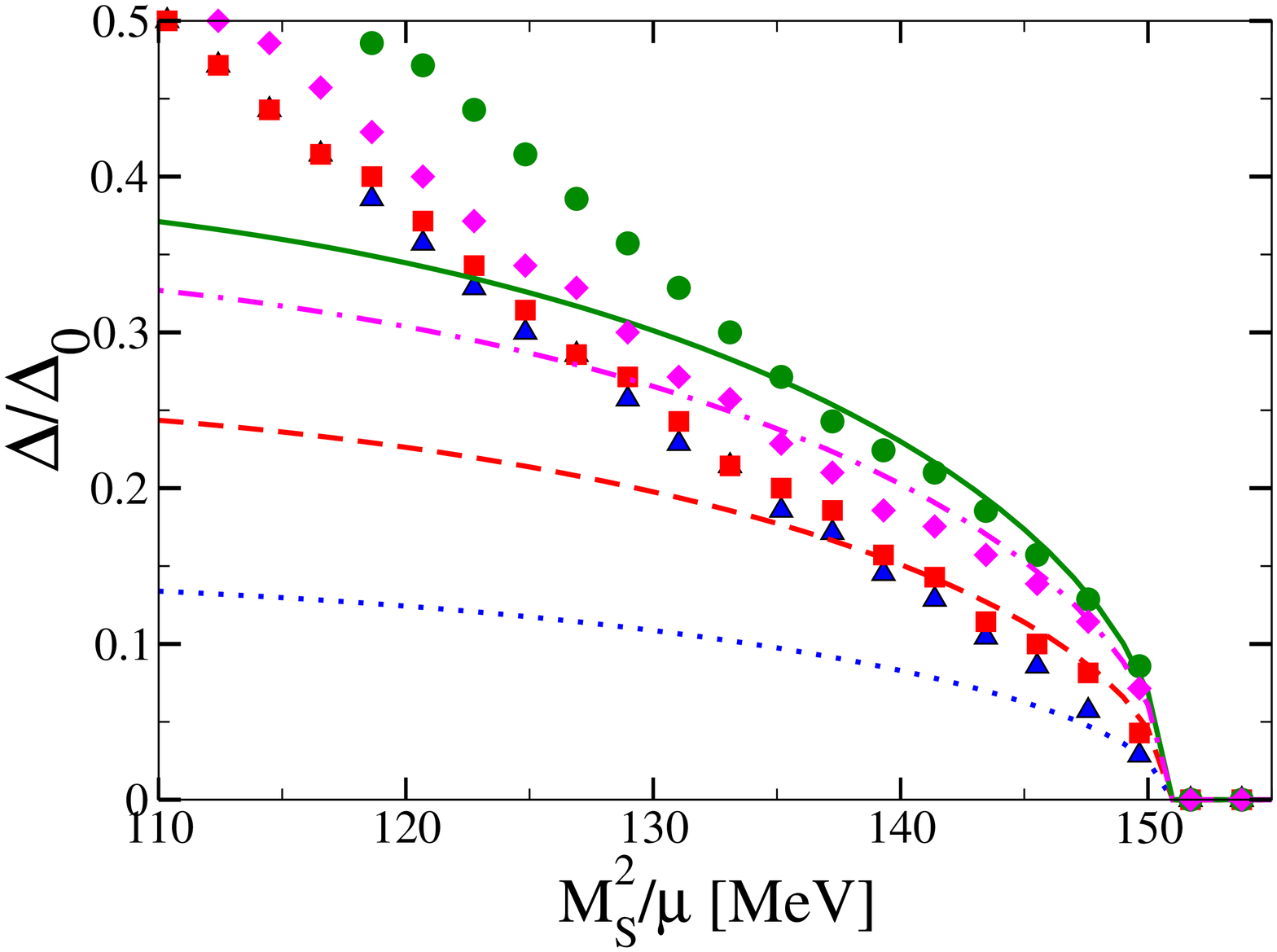}
\includegraphics[width=2.7in,angle=-90]{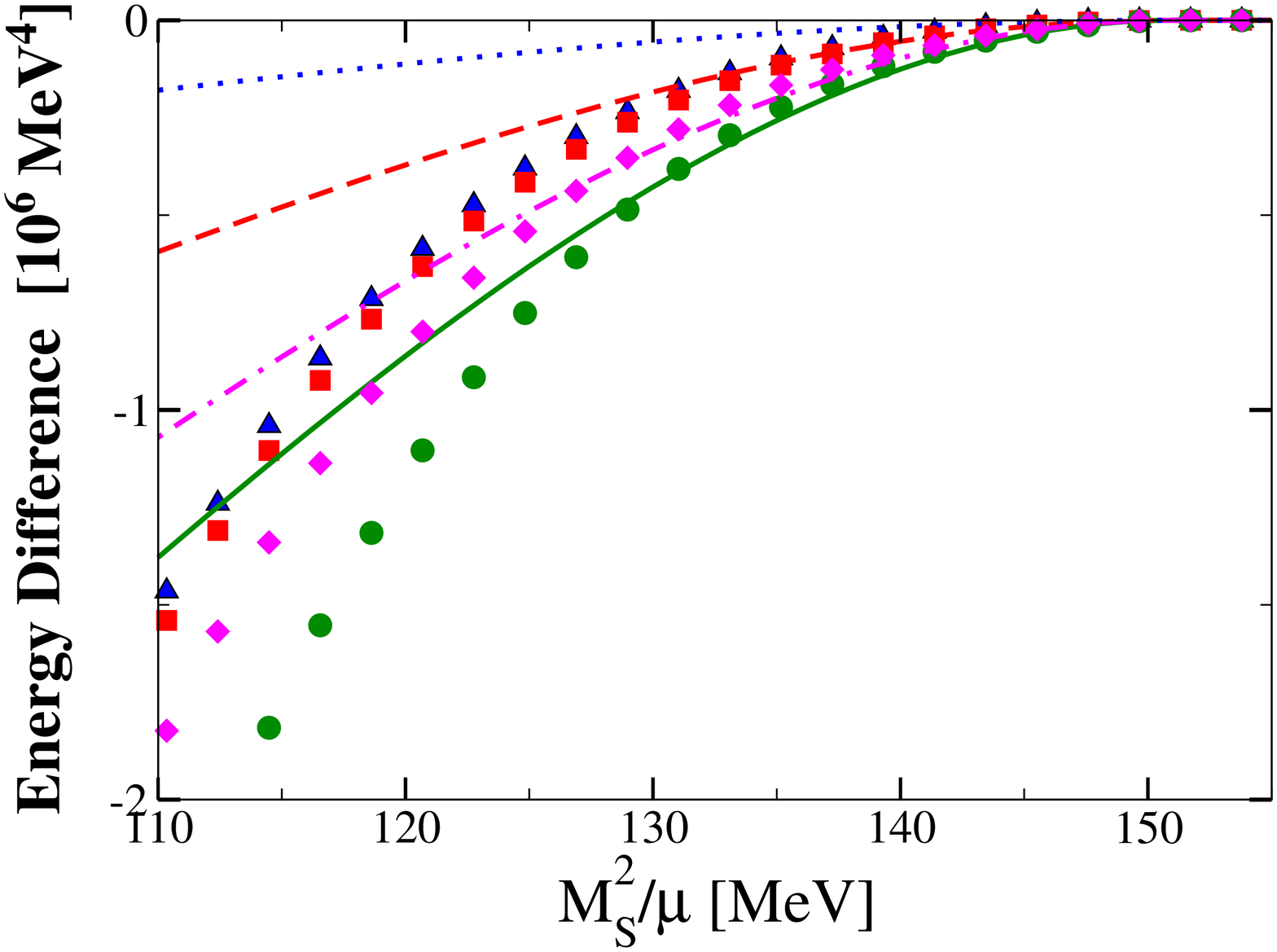}
\caption{Plot of $\Delta/\Delta_0$ (left panel) and of  the free
energy relative to neutral non interacting quark matter (right
panel) as a function of $M_s^2/\mu$ for four  values of the angle
$\phi$ between $\bf q_2$ and $\bf q_3$. The various lines correspond
to the calculations done in the Ginzburg-Landau approximation
described  in Section III  whereas dots correspond to the NJL
calculation of Section IV, done without making a Ginzburg-Landau
approximation. The full lines (green online)  and circles correspond
to $\phi=0$, the dashed-dotted lines (magenta online) and diamonds
correspond to $\phi=2\pi /3$, the dashed lines (red online) and
squares correspond to $\phi=7\pi/8$,  the dotted lines (blue online)
and triangles correspond to $\phi=31\pi/32$.} \label{figgap}
\end{figure*}

Even with a rather small
value of $\Delta$ --- in Fig.~\ref{pairingregions} $\Delta/\delta\mu=0.056$ --- the
ribbons on the Fermi surfaces where pairing occurs do not look very narrow.  In fact,
if we increase $\Delta$ to the point at which $\Delta/\delta\mu=0.19$, keeping all other
parameters as in Fig.~\ref{pairingregions}, the blocking regions at the north and south poles
visible in Fig.~\ref{pairingregions} disappear entirely. The
pairing ``ribbons'' become so wide
that the ribbon centered at $\theta=67.1^\circ$ expands to encompass $\theta=0$,
becoming more of a hat for the Fermi surface than a ribbon on it.
Furthermore, even in Fig.~\ref{pairingregions} where the pairing ribbons
on the Fermi surfaces are somewhat narrow, in that
there {\it are} blocking regions at the poles,
these blocking regions are surrounded by regions of momentum space
where pairing is quite significant. Just a little distance in $\ell_\parallel$
away from the Fermi surfaces, the angular extent of the regions where pairing
is significant grows rapidly, becoming much wider than right at the Fermi surfaces
themselves.
These are all indications that even though the Ginzburg-Landau approximation
is formally controlled by the parameter $\Delta/\delta\mu$, it may break
down quantitatively at rather small values of $\Delta/\delta\mu$. After all, in
the Ginzburg-Landau limit $\Delta/\delta\mu\rightarrow 0$
the pairing is dominated by infinitesimally narrow
ribbons exactly where the shifted Fermi surfaces cross.
Using this as a basis for approximation cannot yield
even a qualitative description of the
physics once $\Delta/\delta\mu\sim 0.2$,  as  even with this small a value
of $\Delta/\delta\mu$
the regions where pairing  is significant are no longer confined to narrow ribbons
but have spread over
considerable regions of the Fermi surfaces.
Indeed, the extent of the bright white regions where
pairing is significant in Fig.~\ref{pairingregions}, in which $\Delta/\delta\mu=0.056$, indicates
that the Ginzburg-Landau approximation may cease to be quantitatively reliable
at values of $\Delta/\delta\mu$ below $0.2$.

\section{Comparisons and conclusions \label{results}}

In Fig. \ref{figgap} we compare our results for the gap parameter and
the free energy in the crystalline color superconducting phase
calculated in the Ginzburg-Landau approximation of Section III with
those obtained without making this approximation in Section IV.
We have done the calculations with $\mu=500$~MeV and a coupling
strength chosen so that $\Delta_0$, the BCS gap in the CFL phase
at $M_s=0$, is $25$~MeV.   We vary $M_s$, but plot quantities
versus $M_s^2/\mu$ because the most important effect of nonzero $M_s$
is the splitting between the $d$, $u$ and $s$ Fermi momenta
given in Eq.~(\ref{pF3}),
controlled by $\delta\mu=M_s^2/(8\mu)$.  We analyze the condensate
(\ref{condensate}) with $\Delta_2=\Delta_3\equiv \Delta$ and
$q_2=q_3\equiv q$.  At each value of $M_s^2/\mu$, we choose
$q=\eta \delta\mu = \eta M_s^2/(8\mu)$, where $\eta=1.1997$.
We fix $\mu_3=\mu_8=0$ and $\mu_e=M_s^2/(4\mu)$, as appropriate
for neutral three-flavor crystalline quark matter with
$\Delta/\delta\mu\ll 1$, for which the Ginzburg-Landau approximation is valid.
We leave investigating the extent to which these chemical potentials may
shift at larger $\Delta$ to future work.
We show our results for four values of the angle between
$\bf q_2$ and $\bf q_3$: $\phi= 0$, $2\pi/3$, $7\pi/8$ and $ 31\pi/32$.
The lines correspond to the Ginzburg-Landau analysis of Section III,
where we have plotted $\Delta$ and $\Omega-\Omega_n$ of
Eqs.~(\ref{DeltaGL}) and (\ref{OmegaGL}), using Eqs.~(\ref{betaresult}), (\ref{beta23result}),
and (\ref{Iresult}) and
using Eq.~(\ref{beta23}) to relate $\alpha$ to $\delta\mu$ and hence to $M_s^2/\mu$.
The points correspond to the NJL calculation of Section IV, where we
have minimized $\Omega$ of Eq.~(\ref{Omega}) with respect to $\Delta$.

We see in Fig.~\ref{figgap} that the NJL calculation has a second
order transition at $M_s^2/\mu \sim 151$~MeV, corresponding to
$\delta\mu\sim 0.754\Delta_0$, for all values of the angle $\phi$.
This result is in agreement with the Ginzburg-Landau calculation, in
which the location of the phase transition depends only on $\alpha$,
which is independent of $\phi$. We then see that near the phase
transition, where $\Delta/\Delta_0$ and hence $\Delta/\delta\mu$ are
small, we find good agreement between the NJL calculation and the
Ginzburg-Landau approximation, as expected. For all values of
$\phi$, as $\Delta/\Delta_0$ increases as $M_s^2/\mu$ is decreased
farther from the transition, we see that both $\Delta$ and
$|\Omega-\Omega_n|$ increase more rapidly with decreasing
$M_s^2/\mu$ than predicted by the Ginzburg-Landau calculation. When
the Ginzburg-Landau approximation breaks down, it does so
conservatively, underpredicting both $\Delta$ and
$|\Omega-\Omega_n|$ for the entire one parameter family of ``crystal
structures'' parameterized by $\phi$. (This behavior also occurs in
the two-flavor model with condensate
(\ref{singleplanewave})~\cite{Bowers:2002xr}.) Furthermore, even
where the Ginzburg-Landau approximation has broken down
quantitatively, it correctly predicts the qualitative feature that
at all values of $M_s^2/\mu$ the most favorable crystal structure is
that with $\phi=0$.  As we saw in the discussion of
Fig.~\ref{ribbonsfigure}, this can be attributed at least
qualitatively to the fact that for ${\bf q}_2\parallel {\bf q}_3$
the two pairing ribbons on the $u$-Fermi surface are farthest apart.
We see that the Ginzburg-Landau approximation is useful as a
qualitative guide even where it has broken down quantitatively.

It is evident from Fig.~\ref{figgap} that the extent of the
regime in which the Ginzburg-Landau approximation is quantitatively reliable
is strongly $\phi$-dependent.  In the best case, which it turns out is
$\phi=0$, the results of the Ginzburg-Landau calculation are in good
agreement with those of the full NJL calculation as long as
$\Delta/\Delta_0 \lesssim 0.25$, corresponding to $\Delta/\delta\mu \lesssim 0.35$.
On the one hand, this looks like a somewhat small value of $\Delta/\delta\mu$.
However, it is remarkable that the Ginzburg-Landau approximation
works so well for this {\it large} a value of $\Delta/\delta\mu$: after all,
we saw in the discussion of Fig.~\ref{pairingregions} in
Section IV that for $\phi=0$  even with  $\Delta/\delta\mu$ only $0.19$
the pairing ``ribbons'' that characterize the Ginzburg-Landau approximation
have broadened into ``pairing hats'' encompassing the north and south poles of the
Fermi surfaces.
For larger $\phi$, the Ginzburg-Landau approximation yields
quantitatively reliable results only for much smaller $\Delta$.
For example, with $\phi=31\pi/32$ we have zoomed in on the region
near the second order phase transition and
seen that the Ginzburg-Landau calculation does give results in quantitative
agreement with the full NJL calculation, but only for $\Delta/\Delta_0 \lesssim 0.04$,
corresponding to $\Delta/\delta\mu \lesssim 0.05$.  Why does the regime
of quantitative validity of the Ginzburg-Landau approximation shrink with
increasing $\phi$?  One possible qualitative explanation is related
to the behavior of the pairing regions displayed for $\phi=0$
in Fig.~\ref{pairingregions}.  We saw in that figure that already at quite
small $\Delta/\delta\mu$, pairing is significant over wide swaths of angle
in momentum space, particularly a little away from the Fermi surfaces.
Since at $\phi=\pi$ the two pairing regions on the $u$ Fermi surface
overlap completely, the closer $\phi$ gets to $\pi$ the smaller the value of
$\Delta/\delta\mu$ needed for the spreading of the pairing regions to become
comparable to the separation between  pairing regions.
This qualitative argument does suggest that the Ginzburg-Landau approximation
should break down at smaller $\Delta/\delta\mu$ for larger $\phi$, as we find.
It does not explain why this breakdown always results in $\Delta$ being enhanced
relative to that predicted by the Ginzburg-Landau approximation.

A quantitative study of the radius of convergence of the Ginzburg-Landau
approximation would require evaluating (at least) the $\Delta^6$ terms,
whose coefficients we shall generically call $\gamma$ .
Because we are working in the vicinity of a point where $\alpha=0$, the
first estimator of the radius of convergence that we can construct comes
by requiring $\gamma \Delta^6 \lesssim (\beta+\beta_{23})\Delta^4$.
The results of the comparison in Fig.~\ref{figgap} are not conclusive on this point,
but they certainly indicate that the radius of convergence in $\Delta$
decreases with increasing $\phi$, decreasing towards zero for
$\phi\rightarrow \pi$.  Given that we see from Fig.~\ref{funcbeta23} that $\beta_{23}$
(and hence $\beta+\beta_{23}$) increases with $\phi$ and diverges
for $\phi\rightarrow\pi$, the apparent behavior of the radius of convergence
in Fig.~\ref{figgap} suggests that $\gamma$ should diverge even faster
for $\phi\rightarrow\pi$.

\begin{figure*}[t]
\includegraphics[width=10cm,angle=-90]{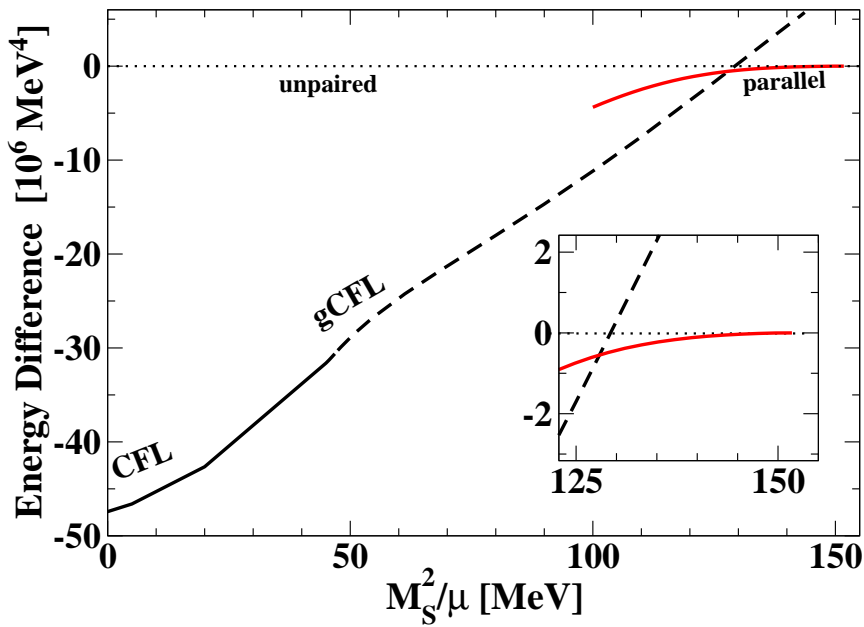}
\caption{Plot  of the free energy of the three-flavor
crystalline color superconducting phase with condensate (\ref{condensate})
with $\Delta_1=0$, $\Delta_2=\Delta_3$, and ${\bf q}_2 \parallel {\bf q}_3$,
compared to the free energy of the CFL and gapless CFL phases.
As throughout,  we are normalizing the interaction to yield
a CFL gap parameter at $M_s=0$ of $\Delta_0=25$~MeV.
We have plotted the free energy calculated as in Section IV, without making
a Ginzburg-Landau approximation.
We see that this crystalline phase, with its particularly simple ``crystal'' structure,
is favored over the other phases shown for
$128~{\rm MeV} < M_s^2/\mu < 151~{\rm MeV}$.
}
 \label{figomega}
\end{figure*}

In Fig. \ref{figomega} we compare our results for the free energy
of the three-flavor crystalline color superconductor with the
most favorable ``crystal'' structure that we have found, namely that with
${\bf q}_2 \parallel {\bf q}_3$, to the free energy of the CFL and
gapless CFL phases~\cite{Alford:2003fq,Alford:2004hz,Fukushima:2004zq}.
Recall that, as discussed in
Section I, the gapless CFL phase
is known to be unstable with respect to counterpropagating
currents~\cite{Casalbuoni:2004tb}.  This means that at every value of $M_s^2/\mu$ where
the CFL phase has been superseded by the
gCFL phase and where the gCFL phase has lower free energy than
unpaired quark matter, there must be one or more phases with
free energy below that of the gCFL phase.  We see that the crystalline
phase we have analyzed, with its particularly simple crystal structure
consisting of only two plane waves, one for the $\langle ud\rangle$ condensate
and one for the $\langle us\rangle$ condensate, has a lower free energy
than both the gCFL and unpaired phases in
a window $128~{\rm MeV}<M_s^2/\mu<151~{\rm MeV}$.
Recalling that the Ginzburg-Landau approximation is reliable over the
widest domain precisely for the case with ${\bf q}_2 \parallel {\bf q}_3$,
we note from Fig.~\ref{figgap} that the Ginzburg-Landau analysis
provides  good approximation to the results of the full NJL calculation
within the window $128~{\rm MeV}<M_s^2/\mu<151~{\rm MeV}$.
Hence, our result for the extent
of this window agrees with that in Ref.~\cite{Casalbuoni:2005zp}, which was obtained
in the Ginzburg-Landau approximation.

We do not actually expect the condensate we have analyzed to be the
stable, lowest free energy, phase.  Instead, we expect that, as in
the investigation of two-flavor crystalline color superconductivity
in Ref.~\cite{Bowers:2002xr}, crystal structures with more plane
waves will yield significantly lower free energy. It will be
interesting to investigate which crystal structures are most
favorable, and over what regime of $M_s^2/(4\mu)$ they have lower
free energy than the gCFL phase. Looking ahead to such an
investigation, the implications of our tests of the Ginzburg-Landau
approximation in the simple case in which we have been able to
calculate both without and with this approximation can be seen as
either a glass half empty or a glass half full.   On the one hand,
we find that the approximation is quantitatively reliable only for
values of $\Delta/\delta\mu$ that are small and for some crystal
structures (those with ${\bf q}_2$ and ${\bf q_3}$ close to
antiparallel) very small. This means that if there are robust
crystal structures with large gaps and condensation energies, they
may not be described reliably within the Ginzburg-Landau
approximation.  On the other hand, we find that even when it breaks
down quantitatively the Ginzburg-Landau approximation remains useful
as a qualitative guide, correctly predicting that the favored
``crystal'' structure among our one parameter family of
possibilities is that with ${\bf q}_2\parallel {\bf q_3}$.
Furthermore, the cases in which the domain of reliability of the
approximation is smallest, those with ${\bf q}_2$ and ${\bf q}_3$
close to antiparallel, are also the cases with the least favorable
free energy. This suggests that when more complicated crystal
structures are analyzed, among them the domain in which the
Ginzburg-Landau approximation is reliable may prove wider for those
structures which are more favorable (i.e. have lower free energy).
As an example, recall that in our analysis we found narrow domains
of reliability for the Ginzburg-Landau approximation for $\phi$ near
$\pi$, the same regime where we found that the condensate we
analyzed was anyway not favored because it had a larger free energy
than one in which only two flavors of quarks pair. Finally, in all
the cases where we have been able to test it, when the
Ginzburg-Landau approximation breaks down it does so conservatively,
underpredicting the magnitude of $\Delta$ and the favorability of
the free energy. We conclude that there is strong motivation to
explore more complicated and realistic crystal structures at least
initially using this approximation.

\vspace{0.8cm}

\noindent
{\bf Acknowledgement} \\
We thank R.~Gatto, E.~Gubankova, G.~Nardulli,  M.~Ruggieri and
A.~Schmitt for useful discussions. The work of MM has been supported
by the ``Bruno Rossi" fellowship program. KR and RS acknowledge the
hospitality of the Nuclear Theory Group at LBNL. This research was
supported in part by the Office of Nuclear Physics of the Office of
Science of the U.S.~Department of Energy under contract
\#DE-AC02-05CH11231 and cooperative research agreement
\#DF-FC02-94ER40818.


\begin{thebibliography}{99}

\bibitem{reviews}
For reviews, see
K.~Rajagopal and F.~Wilczek,
arXiv:hep-ph/0011333;
M.~G.~Alford,
Ann.\ Rev.\ Nucl.\ Part.\ Sci.\  {\bf 51}, 131 (2001)
[arXiv:hep-ph/0102047];
G.~Nardulli,
Riv.\ Nuovo Cim.\  {\bf 25N3}, 1 (2002)
[arXiv:hep-ph/0202037];
S.~Reddy,
Acta Phys.\ Polon.\ B {\bf 33}, 4101 (2002)
[arXiv:nucl-th/0211045];
T.~Sch\"afer,
arXiv:hep-ph/0304281;
D.~H.~Rischke, Prog.\ Part.\ Nucl.\ Phys.\ {\bf 52}, 197 (2004)
[arXiv:nucl-th/0305030];
  M.~Alford,
  Prog.\ Theor.\ Phys.\ Suppl.\  {\bf 153}, 1 (2004)
  [arXiv:nucl-th/0312007];
  M.~Buballa,
  Phys.\ Rept.\  {\bf 407}, 205 (2005)
  [arXiv:hep-ph/0402234];
  H.~c.~Ren,
  arXiv:hep-ph/0404074;
I. Shovkovy, arXiv:nucl-th/0410091;
T. Sch\"afer, arXiv:hep-ph/0509068.


\bibitem{BCS}
J.\ Bardeen, L.N.\ Cooper, and J.R.\ Schrieffer,
Phys.\ Rev.\ {\bf 108}, 1175 (1957).

\bibitem{Iwasaki:1994ij}
  M.~Iwasaki and T.~Iwado,
  Phys.\ Lett.\ B {\bf 350}, 163 (1995).
T.~Sch\"afer,
Phys.\ Rev.\ D {\bf 62}, 094007 (2000)
[arXiv:hep-ph/0006034];
M.~Buballa, J.~Hosek and M.~Oertel,
Phys.\ Rev.\ Lett.\  {\bf 90}, 182002 (2003)
[arXiv:hep-ph/0204275];
A.~Schmitt, Q.~Wang and D.~H.~Rischke,
Phys.\ Rev.\ D {\bf 66}, 114010 (2002)
[arXiv:nucl-th/0209050];
M.~G.~Alford, J.~A.~Bowers, J.~M.~Cheyne and G.~A.~Cowan,
Phys.\ Rev.\ D {\bf 67}, 054018 (2003)
[arXiv:hep-ph/0210106];
  A.~Schmitt,
  Phys.\ Rev.\ D {\bf 71}, 054016 (2005)
  [arXiv:nucl-th/0412033].

\bibitem{Alford:1997zt}
  M.~G.~Alford, K.~Rajagopal and F.~Wilczek,
  Phys.\ Lett.\ B {\bf 422}, 247 (1998)
  [arXiv:hep-ph/9711395].


\bibitem{Alford:1998mk}
M.~G.~Alford, K.~Rajagopal and F.~Wilczek,
Nucl.\ Phys.\ B {\bf 537}, 443 (1999)
[arXiv:hep-ph/9804403].

\bibitem{Alford:2002kj}
M.~Alford and K.~Rajagopal,
JHEP {\bf 0206}, 031 (2002)
[arXiv:hep-ph/0204001].


\bibitem{Iida:2000ha}
K.~Iida and G.~Baym,
Phys.\ Rev.\ D {\bf 63}, 074018 (2001)
[Erratum-ibid.\ D {\bf 66}, 059903 (2002)]
[arXiv:hep-ph/0011229];


\bibitem{Amore:2001uf}
Stable bulk matter must be neutral under
all gauged charges, whether they are spontaneously broken or not.
In the case of the electromagnetic gauge symmetry, this simply
requires zero charge density.
In the case of the color gauge symmetry,
bulk matter must in fact be a color singlet, which is a more
restrictive condition than mere color neutrality.
However, the
free energy cost of projecting a color neutral state onto a
color singlet state
falls rapidly with volume, as
long as we are considering volumes larger than the size of
a Cooper pair.
Given that if quark matter occurs within
the core of a neutron star the relevant volumes will be of order
cubic kilometers, whereas Cooper pairs have sizes of order fm,
it is more than sufficient to consider only the consequences
of enforcing color neutrality. See
P.~Amore, M.~C.~Birse, J.~A.~McGovern and N.~R.~Walet,
Phys.\ Rev.\ D {\bf 65}, 074005 (2002)
[arXiv:hep-ph/0110267].


\bibitem{Steiner:2002gx}
A.~W.~Steiner, S.~Reddy and M.~Prakash,
Phys.\ Rev.\ D {\bf 66}, 094007 (2002)
[arXiv:hep-ph/0205201].


\bibitem{Huang:2002zd}
  M.~Huang, P.~f.~Zhuang and W.~q.~Chao,
  Phys.\ Rev.\ D {\bf 67}, 065015 (2003)
  [arXiv:hep-ph/0207008];
F.~Neumann, M.~Buballa and M.~Oertel,
Nucl.\ Phys.\ A {\bf 714}, 481 (2003)
[arXiv:hep-ph/0210078].




\bibitem{Alford:2003fq}
M.~Alford, C.~Kouvaris and K.~Rajagopal,
Phys.\ Rev.\ Lett.\  {\bf 92}, 222001 (2004)
[arXiv:hep-ph/0311286].

\bibitem{Alford:2004hz}
  M.~Alford, C.~Kouvaris and K.~Rajagopal,
  Phys.\ Rev.\ D {\bf 71}, 054009 (2005)
  [arXiv:hep-ph/0406137].

\bibitem{Alford:2004nf}
  M.~Alford, C.~Kouvaris and K.~Rajagopal,
  arXiv:hep-ph/0407257.

\bibitem{Ruster:2004eg}
S.~B.~R\"uster, I.~A.~Shovkovy and D.~H.~Rischke,
Nucl.\ Phys.\ A {\bf 743}, 127 (2004)
[arXiv:hep-ph/0405170].

\bibitem{Fukushima:2004zq}
  K.~Fukushima, C.~Kouvaris and K.~Rajagopal,
  Phys.\ Rev.\ D {\bf 71}, 034002 (2005)
  [arXiv:hep-ph/0408322].

\bibitem{Alford:2004zr}
  M.~Alford, P.~Jotwani, C.~Kouvaris, J.~Kundu and K.~Rajagopal,
  Phys.\ Rev.\ D {\bf 71}, 114011 (2005)
  [arXiv:astro-ph/0411560].

\bibitem{Abuki:2004zk}
  H.~Abuki, M.~Kitazawa and T.~Kunihiro,
  Phys.\ Lett.\ B {\bf 615}, 102 (2005)
  [arXiv:hep-ph/0412382].

\bibitem{Ruster:2005jc}
  S.~B.~R\"uster, V.~Werth, M.~Buballa, I.~A.~Shovkovy and D.~H.~Rischke,
  Phys.\ Rev.\ D {\bf 72}, 034004 (2005)
  [arXiv:hep-ph/0503184].

\bibitem{Shovkovy:2003uu}
I.~Shovkovy and M.~Huang,
Phys.\ Lett.\ B {\bf 564}, 205 (2003)
[arXiv:hep-ph/0302142];
  M.~Huang and I.~Shovkovy,
  Nucl.\ Phys.\ A {\bf 729}, 835 (2003)
  [arXiv:hep-ph/0307273].

\bibitem{Gubankova:2003uj}
E.~Gubankova, W.~V.~Liu and F.~Wilczek,
Phys.\ Rev.\ Lett.\  {\bf 91}, 032001 (2003).

\bibitem{Huang:2004bg}
M.~Huang and I.~A.~Shovkovy,
Phys. Rev. D {\bf 70}, 051501 (2004)
[arXiv:hep-ph/0407049];
  M.~Huang and I.~A.~Shovkovy,
  Phys.\ Rev.\ D {\bf 70}, 094030 (2004)
  [arXiv:hep-ph/0408268];
  I.~Giannakis and H.~C.~Ren,
  Phys.\ Lett.\ B {\bf 611}, 137 (2005)
  [arXiv:hep-ph/0412015];
  M.~Alford and Q.~h.~Wang,
  J.\ Phys.\ G {\bf 31}, 719 (2005)
  [arXiv:hep-ph/0501078];
  M.~Huang,
  Phys.\ Rev.\ D {\bf 73}, 045007 (2006)
  [arXiv:hep-ph/0504235].


\bibitem{Casalbuoni:2004tb}
  R.~Casalbuoni, R.~Gatto, M.~Mannarelli, G.~Nardulli and M.~Ruggieri,
  Phys.\ Lett.\ B {\bf 605}, 362 (2005)
  [Erratum-ibid.\ B {\bf 615}, 297 (2005)]
  [arXiv:hep-ph/0410401];
  K.~Fukushima,
  Phys.\ Rev.\ D {\bf 72}, 074002 (2005)
  [arXiv:hep-ph/0506080].



\bibitem{Bedaque:2003hi}
  P.~F.~Bedaque, H.~Caldas and G.~Rupak,
  Phys.\ Rev.\ Lett.\  {\bf 91}, 247002 (2003)
  [arXiv:cond-mat/0306694];
  H.~Caldas,
  Phys.\ Rev.\ A {\bf 69}, 063602 (2004)
  [arXiv:hep-ph/0312275];
  M.~M.~Forbes,  E.~Gubankova, W.~V.~Liu and F.~Wilczek,
  Phys.\ Rev.\ Lett.\  {\bf 94}, 017001 (2005)
  [arXiv:hep-ph/0405059];
  J.~Carlson and S.~Reddy,
  Phys.\ Rev.\ Lett.\  {\bf 95}, 060401 (2005)
  [arXiv:cond-mat/0503256].



\bibitem{KetterleImbalancedSpin} M.W.~Zwierlein,
A.~Schirotzek, C.H.~Schunck, and W.~Ketterle, Science {\bf 311},
492 (2006) [arXiv:cond-mat/0511197].

\bibitem{HuletPhaseSeparation} G.B.~Partridge, W.~Li, R.I.~Kamar,
Y.-a.~Liao, and R.G.~Hulet,
 Science {\bf 311}, 503 (2006) [arXiv:cond-mat/0511752].



\bibitem{Bailin:1983bm}
  D.~Bailin and A.~Love,
  Phys.\ Rept.\  {\bf 107}, 325 (1984).

\bibitem{Rapp:1997zu}
  R.~Rapp, T.~Sch\"afer, E.~V.~Shuryak and M.~Velkovsky,
  Phys.\ Rev.\ Lett.\  {\bf 81}, 53 (1998)
  [arXiv:hep-ph/9711396].

\bibitem{Rajagopal:2005dg}
  K.~Rajagopal and A.~Schmitt,
  Phys.\ Rev.\ D {\bf 73}, 045003 (2006)
  [arXiv:hep-ph/0512043].

\bibitem{Alford:2000ze}
M.~G.~Alford, J.~A.~Bowers and K.~Rajagopal,
Phys.\ Rev.\ D {\bf 63}, 074016 (2001)
[arXiv:hep-ph/0008208];

\bibitem{Bowers:2001ip}
J.~.A.~Bowers, J.~Kundu, K.~Rajagopal and E.~Shuster,
Phys.\ Ref.\ D {\bf 64}, 014024 (2001)
[arXiv:hep-ph/0101067];


\bibitem{Casalbuoni:2001gt}
  R.~Casalbuoni, R.~Gatto, M.~Mannarelli and G.~Nardulli,
  Phys.\ Lett.\ B {\bf 511}, 218 (2001)
  [arXiv:hep-ph/0101326].

\bibitem{Leibovich:2001xr}
A.~K.~Leibovich, K.~Rajagopal and E.~Shuster, Phys.\ Rev.\ D {\bf 64},
094005 (2001) [arXiv:hep-ph/0104073];

\bibitem{Kundu:2001tt}
J.~Kundu and K.~Rajagopal,
Phys.\ Rev.\ D {\bf 65}, 094022 (2002)
[arXiv:hep-ph/0112206];

\bibitem{Casalbuoni:2003wh}
R.~Casalbuoni and G.~Nardulli,
Rev.\ Mod.\ Phys.\  {\bf 263}, 320 (2004)
[arXiv:hep-ph/0305069];


\bibitem{Casalbuoni:2004wm}
  R.~Casalbuoni, M.~Ciminale, M.~Mannarelli, G.~Nardulli, M.~Ruggieri and
R.~Gatto,
  Phys.\ Rev.\ D {\bf 70}, 054004 (2004)
  [arXiv:hep-ph/0404090].




\bibitem{Bowers:2002xr}
J.~A.~Bowers and K.~Rajagopal,
Phys.\ Rev.\ D {\bf 66}, 065002 (2002)
[arXiv:hep-ph/0204079].

\bibitem{Casalbuoni:2005zp}
  R.~Casalbuoni, R.~Gatto, N.~Ippolito, G.~Nardulli and M.~Ruggieri,
  Phys.\ Lett.\ B {\bf 627}, 89 (2005)
  [arXiv:hep-ph/0507247].

\bibitem{Ciminale:2006sm}
  M.~Ciminale, G.~Nardulli, M.~Ruggieri and R.~Gatto,
  arXiv:hep-ph/0602180.

\bibitem{LOFF}
A.~I.~Larkin and Yu.~N.~Ovchinnikov, Zh. Eksp. Teor. Fiz.~{\bf 47}, 1136
(1964)[Sov. Phys. JETP {\bf 20}, 762 (1965)];
P.~Fulde and R.~A.~Ferrell, Phys.\ Rev.\ {\bf 135}, A550 (1964);
S.~Takada and T.~Izuyama, Prog.\  Theor.\ Phys.\ {\bf 41}, 635 (1969).


\bibitem{Bedaque:2001je}
P.~F.~Bedaque and T.~Sch\"afer,
Nucl.\ Phys.\ A {\bf 697}, 802 (2002)
[arXiv:hep-ph/0105150];
D.~B.~Kaplan and S.~Reddy,
Phys.\ Rev.\ D {\bf 65}, 054042 (2002)
[arXiv:hep-ph/0107265];
  A.~Kryjevski, D.~B.~Kaplan and T.~Sch\"afer,
  Phys.\ Rev.\ D {\bf 71}, 034004 (2005)
  [arXiv:hep-ph/0404290];

\bibitem{Kryjevski:2004jw}
  A.~Kryjevski and T.~Sch\"afer,
  Phys.\ Lett.\ B {\bf 606}, 52 (2005)
  [arXiv:hep-ph/0407329];
  A.~Kryjevski and D.~Yamada,
  Phys.\ Rev.\ D {\bf 71}, 014011 (2005)
  [arXiv:hep-ph/0407350].
  M.~Buballa,
  Phys.\ Lett.\ B {\bf 609}, 57 (2005)
  [arXiv:hep-ph/0410397];
  M.~M.~Forbes,
  Phys.\ Rev.\ D {\bf 72}, 094032 (2005)
  [arXiv:hep-ph/0411001].



\bibitem{Kryjevski:2005qq}
  A.~Kryjevski,
  arXiv:hep-ph/0508180;
  T.~Schafer,
  Phys.\ Rev.\ Lett.\  {\bf 96}, 012305 (2006)
  [arXiv:hep-ph/0508190];
see also
  D.~K.~Hong,
  arXiv:hep-ph/0506097.

\bibitem{Hong:1998tn}
  D.~K.~Hong,
  Phys.\ Lett.\ B {\bf 473}, 118 (2000)
  [arXiv:hep-ph/9812510];
  D.~K.~Hong,
  Nucl.\ Phys.\ B {\bf 582}, 451 (2000)
  [arXiv:hep-ph/9905523];
  S.~R.~Beane, P.~F.~Bedaque and M.~J.~Savage,
  Phys.\ Lett.\ B {\bf 483}, 131 (2000)
  [arXiv:hep-ph/0002209];
  R.~Casalbuoni, R.~Gatto and G.~Nardulli,
  Phys.\ Lett.\ B {\bf 498}, 179 (2001)
  [Erratum-ibid.\ B {\bf 517}, 483 (2001)]
  [arXiv:hep-ph/0010321];
  R.~Casalbuoni, R.~Gatto, M.~Mannarelli and G.~Nardulli,
  Phys.\ Lett.\ B {\bf 524}, 144 (2002)
  [arXiv:hep-ph/0107024];
  T.~Schafer,
  Phys.\ Rev.\ D {\bf 65}, 074006 (2002)
  [arXiv:hep-ph/0109052].
  T.~Schafer,
  Nucl.\ Phys.\ A {\bf 728}, 251 (2003)
  [arXiv:hep-ph/0307074].

\bibitem{Nardulli:2002ma}
G. Nardulli, in Ref.~\cite{reviews}.


\bibitem{Gerhold:2003js}
A.~Gerhold and A.~Rebhan,
Phys.\ Rev.\ D {\bf 68}, 011502 (2003)
[arXiv:hep-ph/0305108];
A.~Kryjevski,
Phys.\ Rev.\ D {\bf 68}, 074008 (2003)
[arXiv:hep-ph/0305173];
  A.~Gerhold,
  Phys.\ Rev.\ D {\bf 71}, 014039 (2005)
  [arXiv:hep-ph/0411086].
D.~D.~Dietrich and D.~H.~Rischke,
Prog.\ Part.\ Nucl.\ Phys.\  {\bf 53}, 305 (2004)
[arXiv:nucl-th/0312044];

\bibitem{Rajagopal:2000ff}
K.~Rajagopal and F.~Wilczek,
Phys.\ Rev.\ Lett.\  {\bf 86}, 3492 (2001)
[arXiv:hep-ph/0012039].


\end{thebibliography}
\end{document}